\documentclass[paper=letter, fontsize=12pt]{article}
\usepackage[english]{babel} 
\usepackage{amsmath,amsfonts,amsthm} 
\usepackage[utf8]{inputenc}
\usepackage{float}
\usepackage{lipsum} 
\usepackage{blindtext}
\usepackage{graphicx} 
\usepackage{caption}
\usepackage[sc]{mathpazo} 
\usepackage[T1]{fontenc} 
\usepackage{microtype} 
\usepackage[hmarginratio=1:1,top=32mm,columnsep=20pt]{geometry} 
\usepackage{multicol} 
\usepackage{booktabs} 
\usepackage{float} 
\usepackage{hyperref} 
\usepackage{lettrine} 
\usepackage{paralist} 
\usepackage{abstract} 
\usepackage{titlesec} 
\usepackage{natbib, url}
\usepackage{subfig}
\renewcommand\thesection{\Roman{section}} 
\renewcommand\thesubsection{\Roman{subsection}} 

\titleformat{\section}[block]{\large\scshape\centering}{\thesection.}{1em}{} 
\titleformat{\subsection}[block]{\large}{\thesubsection.}{1em}{} 
\usepackage{fancyhdr} 
\usepackage[bottom]{footmisc}

\newcommand{\given}{\,|\,}

\newcommand{\calA}{\mathcal{A}}
\newcommand{\ba}{\mathbf{a}}

\newcommand{\bs}{\mathbf{s}}

\newcommand{\calS}{\mathcal{S}}
\newcommand{\bC}{\mathbf{C}}
\newcommand{\bY}{\mathbf{Y}}
\newcommand{\by}{\mathbf{y}}
\newcommand{\bX}{\mathbf{X}}
\newcommand{\bx}{\mathbf{x}}
\newcommand{\bZ}{\mathbf{Z}}
\newcommand{\bz}{\mathbf{z}}

\newcommand{\bV}{\mathbf{V}}
\newcommand{\bA}{\mathbf{A}}
\newcommand{\bB}{\mathbf{B}}

\newcommand{\bO}{\mathbf{O}}
\newcommand{\bu}{\mathbf{u}}
\newcommand{\bU}{\mathbf{U}}
\newcommand{\calU}{{\cal U}}

\newcommand{\bM}{\mathbf{M}}

\newcommand{\bI}{\mathbf{I}}
\newcommand{\bD}{\mathbf{D}}
\newcommand{\calD}{\mathcal{D}}
\newcommand{\bL}{\mathbf{L}}
\newcommand{\bR}{\mathbf{R}}
\newcommand{\bK}{\mathbf{K}}
\newcommand{\bv}{\mathbf{v}}
\newcommand{\bE}{\mathbf{E}}
\newcommand{\bP}{\mathbf{P}}
\newcommand{\bmu}{\boldsymbol{\mu}}
\newcommand{\bSigma}{\boldsymbol{\Sigma}}
\newcommand{\beps}{\boldsymbol{\epsilon}}
\newcommand{\bbeta}{\boldsymbol{\beta}}

\newcommand{\bomega}{\boldsymbol{\omega}}
\newcommand{\bgamma}{\boldsymbol{\gamma}}
\newcommand{\etab}{\boldsymbol{\eta}}
\newcommand{\bPsi}{\boldsymbol{\Psi}}
\newcommand{\brho}{\boldsymbol{\rho}}
\newcommand{\bkappa}{\boldsymbol{\mathcal{K}}}

\graphicspath{{./pics/}}

\title{\vspace{-15mm}\fontsize{24pt}{10pt}\selectfont\textbf{High-dimensional Multivariate Geostatistics: A Bayesian Matrix-Normal Approach}} 

\author{
\large
{\textsc{Lu Zhang}}\\[2mm]
{\textsc{UCLA Department of Biostatistics }}\\[2mm]
\normalsize \href{mailto:lu.zhang@ucla.edu}{Lu.Zhang@ucla.edu}\\[2mm] 
\large
{\textsc{Sudipto Banerjee}}\\[2mm]
{\textsc{UCLA Department of Biostatistics }}\\[2mm]
\normalsize \href{mailto:sudipto@ucla.edu}{sudipto@ucla.edu}\\[2mm] 
\large
{\textsc{Andrew O. Finley}}\\[2mm]
{\textsc{Michigan State University Departments of Forestry and Geography}}\\[2mm]
\normalsize \href{mailto:finleya@msu.edu}{finleya@msu.edu}\\[2mm]
}
\date{Dec 06, 2020}
\providecommand{\keywords}[1]{\textbf{\textit{Key words:}} #1}
\begin{document}
\maketitle 
\thispagestyle{fancy} 

\label{firstpage}

\begin{abstract}
Joint modeling of spatially-oriented dependent variables 
{is} commonplace in the environmental sciences, where scientists seek to estimate the relationships among a set of environmental outcomes accounting for dependence among these outcomes and the spatial dependence for each outcome. Such modeling is now sought for 
{massive} data sets with variables measured at a very large number of locations. Bayesian inference, while attractive for accommodating uncertainties through 
hierarchical structures, can become computationally onerous for modeling massive spatial data sets because of 
{its} reliance on iterative estimation algorithms. This manuscript develops a conjugate Bayesian framework for analyzing multivariate spatial data using analytically tractable posterior distributions that obviate iterative algorithms. We discuss differences between modeling the multivariate response itself as a spatial process and that of modeling a latent process in a hierarchical model. We illustrate the computational and inferential benefits of these models using simulation studies and analysis of a Vegetation Index data set with spatially dependent observations numbering in the millions.
\end{abstract}


\newpage
\section{Introduction}\label{sec: Intro}
Analyzing environmental data sets often require joint modeling of multiple spatially dependent variables accounting for dependence among the variables and the spatial association for each variable. 
{Joint modeling approaches have two primary benefits over independently analyzing each variable: (i) we can estimate posited associations among the variables that are precluded by independent analyses; and (ii) we can obtain improved predictive inference by borrowing strength across the variables using their dependence. A further consequence of these benefits is that the estimated residuals from joint or multivariate regression models are a better reflection of random noise as the model accounts for different sources of dependence. 
}
For point-referenced variables, multivariate Gaussian processes (GPs) serve as versatile tools for joint modeling of spatial variables \citep[see, e.g.,][and references therein]{scha04, cressie2015statistics, banerjee2014hierarchical, genton2015cross, wackernagel2003}. These texts discuss the substantial literature on modeling multivariate spatial processes, a field referred to as \emph{multivariate geostatistics}, and include several references that have cogently demonstrated some of the aforementioned benefits of joint modeling over independent modeling. 

However, for a data set with $n$ observed locations, fitting a GP based spatial model typically requires floating point operations (flops) and memory requirements of the order $\sim n^3$ and $\sim n^2$, respectively. This is challenging when $n$ is large. This ``Big Data'' problem in spatial statistics continues to receive much attention and a comprehensive review is beyond the scope of this article \citep[see, e.g.,][]{banerjee2017high, heaton2019case, sun2012geostatistics}. Much of the aforementioned literature for scalable models focused on univariate spatial processes, i.e., assuming only one response for each location. Recently \cite{bradley2015multivariate} and \cite{bradley2018computationally} proposed novel classes of multivariate spatiotemporal models for high-dimensional areal data. On the other hand, our current work emphasizes on multivariate continuous spatial processes as is customarily used for point-referenced geostatistical data analysis.     

Multivariate processes \citep[see, e.g.,][and references therein]{genton2015cross, salvana2020nonstationary, le2006statistical}, has received relatively limited developments in the context of massive data. Bayesian models are attractive for inference on multivariate spatial processes because they can accommodate uncertainties in the process parameters more flexibly through their hierarchical structure. Multivariate spatial interpolation using conjugate Bayesian modeling can be found in \citet[][]{brown1994multivariate,le1997bayesian,sun1998assessment, le2001spatial,gamerman2004multivariate}, but these methods do not address the challenges encountered in massive data sets. More flexible methods for joint modeling, including spatial factor models, have been investigated in Bayesian contexts \citep[see, e.g. ][]{schmidt2003bayesian, renbanerjee2013, taylor2019spatial}, but these methods have focused upon delivering full Bayesian inference through iterative algorithms such as Markov chain Monte Carlo (MCMC), Integrated Nested Laplace Approximations or INLA \citep{ruemartinochopin2009} or variational Bayes \citep{renbanerjee2011vb}.

Our current contribution is 
{to extend conjugate Bayesian multivariate regression models to spatial process settings and deliver inference using exact posterior samples obtained directly from closed-form posterior distributions. We specifically address the scenario where the number of locations is massive but the number of dependent variables is modestly small so that dimension reduction is sought on the number of locations and not on the number of variables. Our primary contribution is, therefore, to see how the conjugate Bayesian multivariate regression models can be adapted to accommodate closed-form posterior distributions and avoid iterative algorithms (such as MCMC or INLA or variational Bayes).} We develop an augmented Bayesian multivariate linear model framework that accommodates conjugate distribution theory, similar to \cite{gamerman2004multivariate}, but that can scale up to massive data sets with locations numbering in the millions. 
{We also extend the univariate conjugate Nearest-Neighbor Gaussian process (NNGP) models in \citet{finley2019efficient} and \citet{zdb2019} to multivariate conjugate spatial regression models. We achieve this by embedding the NNGP covariance matrix as a parameter in the Matrix-Normal and Inverse-Wishart family of conjugate priors for multivariate Bayesian regression. We will consider two classes of models. The first is obtained by modeling the spatially dependent variables jointly as a multivariate spatial process, while the second models a latent multivariate spatial process in a hierarchical setup. We refer to the former as the ``response'' model and the latter as the ``latent'' model. While the univariate versions of the response and latent models have been discussed in \citet{finley2019efficient} and \citet{zdb2019}, here we provide some new theoretical insights that help understand the performance of these models as approximations to full Gaussian processes.}



The balance of our paper is arranged as follows. Section~\ref{sec: spatial modeling} develops a conjugate Bayesian multivariate spatial regression framework using Matrix-Normal and Inverse-Wishart prior distributions. Section~\ref{subsec: conj_models} develops two classes of conjugate multivariate models the response models and latent models and show how they provide posterior distributions in closed forms. Subsequently, in Section~\ref{subsec: scalable_conj_models} we develop scalable versions of these models using the Nearest Neighbor Gaussian process (NNGP) 
and a cross-validation algorithm to fix certain hyperparameters required for closed form posteriors is presented in Section~\ref{subsec: CV_conj_NNGP}. 
Section~\ref{sec: simulation} presents simulation experiments, while Section~\ref{sec: real_data_analy} analyzes a massive Normalized Difference Vegetation Index data with a few million locations. Finally, Section~\ref{sec: Conclusions} concludes the manuscript with some discussion.    

\section{Bayesian Multivariate Geostatistical Modeling}\label{sec: spatial modeling}
\subsection{Conjugate Multivariate Spatial Models}\label{subsec: conj_models}
\paragraph{Conjugate Multivariate Response Model}\label{sec: conj_resp}
Let $\by(\bs) = (y_1(\bs), \ldots, y_q(\bs))^\top \in \mathbb{R}^q$ be a $q\times 1$ vector of outcomes at location $\bs \in \mathcal{D} \subset \mathbb{R}^d$ and let $\bx(\bs) = (x_1(\bs), \ldots, x_p(\bs))^\top \in \mathbb{R}^p$ be a $p\times 1$ vector of explanatory variables observed at $\bs$. Conditional on these explanatory variables, $\by(\bs)$ is modeled as a multivariate Gaussian process,
\begin{equation}\label{eq: spatial_GP_model_proportion}
\by(\bs) \sim \mbox{GP}(\bbeta^\top \bx(\bs), \bC(\cdot, \cdot))\;;\quad \bC(\bs, \bs') = [\rho_{\psi}(\bs, \bs') + (\alpha^{-1} - 1)\delta_{\bs = \bs'}]\bSigma\;,
\end{equation}
where the mean of $\by(\bs)$ is $\bbeta^{\top}\bx(\bs)$, $\bbeta$ is a $p \times q$ matrix of regression coefficients and $\bC(\bs,\bs') = \{\mbox{cov}\{y_i(\bs),y_j(\bs')\}\}$ is a $q\times q$ cross-covariance matrix \citep[][]{genton2015cross} whose $(i,j)$-th element is the covariance between $y_i(\bs)$ and $y_j(\bs')$. The cross-covariance matrix is defined for each pair of locations and is further specified as a multiple of a nonspatial positive definite matrix $\bSigma$. The multiplication factor is a function of the two locations and is composed of two components: a spatial correlation function, $\rho_{\psi}(\bs, \bs')$, which introduces spatial dependence between the outcomes through hyperparameters $\psi$, and a micro-scale adjustment $(1/\alpha - 1)\delta_{\bs = \bs'}$, where $\delta_{\bs=\bs'} = 1$ if $\bs=\bs'$ and $\delta_{\bs=\bs'}=0$ if $\bs\neq\bs'$, and $\alpha\in (0,1]$ is a scalar parameter representing the overall strength of the spatial variability as a proportion of the total variation. 

The covariance among the elements of $\by(\bs)$ within a location $\bs$ is given by the elements of $\bC(\bs,\bs) = (1/\alpha)\bSigma$. Thus, $\bSigma$ is the within-location (nonspatial) dependence among the outcomes adjusted by a scale of $1/\alpha$ to accommodate additional variation at local scales. The interpretation of $\alpha$ is analogous to the ratio of the ``partial sill'' to the ``sill'' in classical geostatistics. For example, in the special case when $\bSigma = \sigma^2\bI_q$, $\mbox{cov}\{y_i(\bs),y_j(\bs')\} = \sigma^2\rho(\bs,\bs') + \sigma^2(1/\alpha - 1)\delta_{\bs=\bs'}$, which shows that $\sigma^2(1/\alpha-1) = \tau^2$ is the variance of micro-scale processes (or the ``nuggets''), so that $\alpha = \sigma^2/(\sigma^2 + \tau^2)$ is the ratio of the spatial variance (partial sill) to the total variance (sill). A similar interpretation for $\alpha$ results in the univariate setting with $q=1$.   

Let ${\cal S} = \{\bs_1, \ldots,  \bs_n\} \subset \mathcal{D}$ be a set of $n$ locations yielding observations on $\by(\bs)$. Then $\bY = \by(\mathcal{S}) = [\by(\bs_1): \cdots : \by(\bs_n)]^\top$ is $n \times q$ and $\bX = \bx(\mathcal{S}) = [\bx(\bs_1) : \cdots : \bx(\bs_n)]^\top$ is the corresponding $n\times p$ matrix of explanatory variables observed over ${\cal S}$. We will assume that $\bX$ has full column rank ($=p < n$). The likelihood emerging from (\ref{eq: spatial_GP_model_proportion}) is $\bY \given \bbeta, \bSigma \sim \mbox{MN}_{n, q}(\bX\bbeta, \bkappa,  \bSigma)$, where $\mbox{MN}$ denotes the Matrix-Normal distribution, $\bkappa = \brho_{\psi} + (\alpha^{-1} - 1)\bI_n$ 
{with $\bI_n$ the $n \times n$ identity matrix and $\brho_{\psi} = \{\rho_{\psi}(\bs_i,\bs_j)\}$ the $n \times n$ spatial correlation matrix. A random matrix $\mathbf{Z}_{n \times p}$ that follows a Matrix-Normal distribution $\mbox{MN}_{n,p}(\mathbf{M}, \bU, \bV)$ has a probability density function 
\begin{equation}\label{eq: MN_density}
p(\bZ\mid\bM, \bU, \bV) = \frac{\exp\left( -\frac{1}{2} \, \mbox{tr}\left[ \bV^{-1} (\bZ - \bM)^{T} \bU^{-1} (\bZ - \bM) \right] \right)}{(2\pi)^{np/2} |\bV|^{n/2} |\bU|^{p/2}}\; ,
\end{equation}
where $\mbox{tr}$ denotes trace, $\bM$ is the mean matrix, $\bU$ is the first scale matrix with dimension $n \times n$ and $\bV$ is the second scale matrix with dimension $p \times p$ \citep{ding2014dimension}. The vectorization of the $n \times p$ random matrix $\bZ = [\bz_1, \ldots, \bz_p]$, denoted $\mbox{vec}(\bZ) = [\bz_1^\top, \ldots, \bz_p^\top]^\top$, follows a Gaussian distribution
$\mbox{vec}(\bZ) \sim \mbox{N}_{np}(\mbox{vec}(\bM), \bV \otimes \bU)$, where $\otimes$ denote the Kronecker product.} 
A conjugate Bayesian model is obtained by a Matrix-Normal-Inverse-Wishart (MNIW) prior on $\{\bbeta, \bSigma\}$, which we denote as
\begin{equation}\label{eq: MNIW prior}
   \mbox{MNIW}(\bbeta,\bSigma\given \bmu_{\beta}, \bV_{r}, \bPsi, \nu) = \mbox{IW}(\bSigma\given \bPsi, \nu)\times \mbox{MN}_{p, q}(\bbeta\given \bmu_{\bbeta} , \bV_r, \bSigma)\; ,
\end{equation}
where $\mbox{IW}(\bSigma\given \cdot,\cdot)$ is the inverted-Wishart distribution. The MNIW family is a conjugate prior with respect to the likelihood (\ref{eq: MN_density}) and, for any fixed values of $\alpha$, $\psi$ and the hyperparameters in the prior density, we obtain the posterior density
\begin{equation}\label{eq: MNIW_posterior}
 \begin{split}
 p(\bbeta,\bSigma\given \bY) &\propto \mbox{MNIW}(\bbeta,\bSigma\given \bmu_{\beta}, \bV_{r}, \bPsi, \nu) \times \mbox{MN}_{n,q}(\bY\given\bX\bbeta, \bkappa, \bSigma) \propto \mbox{MNIW}(\bmu^\ast, \bV^\ast, \mathbf{\Psi}^\ast, \nu^\ast)\;, 
 \end{split}
\end{equation}
where
\begin{equation}\label{eq: collapsed_spatial_pars1}
\begin{aligned}
    \bV^\ast &= (\bX^\top \bkappa^{-1} \bX + \bV_r^{-1})^{-1}\; , \; \bmu^\ast = \bV^\ast (\bX^\top \bkappa^{-1}\bY + \bV_r^{-1}\bmu_{\bbeta})\; ,\\
    \bPsi^\ast &= \bPsi + \bY^\top \bkappa^{-1}\bY
    +\bmu_{\bbeta}^\top \bV_r^{-1} \bmu_{\bbeta} - \bmu^{\ast\top} \bV^{\ast-1} \bmu^\ast\; , \mbox{ and  }\nu^\ast = \nu + n\;.
\end{aligned}  
\end{equation}
Direct sampling from the $\mbox{MNIW}$ posterior distribution in (\ref{eq: MNIW_posterior}) is achieved by first sampling $\bSigma \sim \mbox{IW}(\mathbf{\Psi}^\ast, \nu^\ast)$ and then sampling one draw of $\bbeta \sim \mbox{MN}_{p,q}(\bmu^{\ast}, \bV^{\ast}, \bSigma)$ for each draw of $\bSigma$. The resulting pairs $\{\bbeta,\bSigma\}$ will be samples from (\ref{eq: MNIW_posterior}). Since this scheme draws directly from the posterior distribution, the sample is exact and does not require burn-in or convergence. 

Turning to predictions, let ${\cal U} = \{\bu_1, \ldots, \bu_{n'}\}$ be a finite set of locations where we intend to predict or impute the value of $\by(\bs)$ based upon an observed $n'\times p$ design matrix $\bX_\calU = [\bx(\bu_1) : \cdots : \bx(\bu_{n'})]^\top$ for ${\cal U}$. If $\bY_\calU = [\by(\bu_1): \cdots : \by(\bu_{n'})]^\top$ is the $n'\times q$ matrix of predictive random variables, then the conditional predictive distribution  is
\begin{equation}\label{eq: posterior_predictive_conditional}
\begin{aligned}
p(\bY_\calU \given \bY, \bbeta, \bSigma) =\mbox{MN}_{n',q}(\bY_\calU\given \bX_\calU \bbeta &+ \brho_{\psi}(\calU, \calS)\bkappa^{-1}[\bY - \bX\bbeta], \\
& \quad \brho_{\psi}(\calU, \calU) + (\alpha^{-1} - 1)\bI_{n'} - \brho_{\psi}(\calU, \calS)\bkappa^{-1}\brho_{\psi}(\calS, \calU),\; \bSigma) \;,
\end{aligned}
\end{equation}
where $\brho_{\psi}(\calU, \calS) = \{\rho_{\psi}(\bu_i,\bs_j)\}$ is $n'\times n$ and $\brho_{\psi}(\calS, \calU) = \{\rho_{\psi}(\bs_i,\bu_j)\} = \brho_{\psi}(\calU, \calS)^{\top}$. Predictions can be directly carried out in posterior predictive fashion, where we sample from
\begin{multline}\label{eq: posterior_predictive}
p(\bY_\calU \given \bY) = \int \mbox{MN}_{n',q}(\bX_\calU \bbeta + \brho_{\psi}(\calU, \calS)\bkappa^{-1}[\bY - \bX\bbeta],\\
\brho_{\psi}(\calU, \calU) + (\alpha^{-1} - 1)\bI_{n'} - \brho_{\psi}(\calU, \calS)\bkappa^{-1}\brho_{\psi}(\calS, \calU),\; \bSigma)
 \times \mbox{MNIW}(\bmu^\ast, \bV^\ast, \mathbf{\Psi}^\ast, \nu^\ast)\,d\bbeta\,d\bSigma\;.
\end{multline}
Sampling from (\ref{eq: posterior_predictive}) is achieved by drawing one $\bY_{\calU}$ from (\ref{eq: posterior_predictive_conditional}) for each posterior draw of $\{\bbeta,\bSigma\}$.  

\paragraph{Conjugate Multivariate Latent Model}\label{subsec: con_latent}
We now discuss a conjugate Bayesian model for a latent process. Consider the spatial regression model 
\begin{equation}\label{eq: conj_latent_model}
    \by(\bs) = \bbeta^\top\bx(\bs) + \bomega(\bs) + \beps(\bs)\;, \; \bs \in {\cal D}\;,
\end{equation}
where $\bomega(\bs) \sim \mbox{GP}(\mathbf{0}_{q \times 1}, \rho_{\psi}(\cdot, \cdot)\bSigma)$ is a $q\times 1$ multivariate latent process with cross-covariance matrix $\rho_{\psi}(\bs,\bs')\bSigma$ and $\beps(\bs) \stackrel{iid}{\sim}\mbox{N}(\mathbf{0}_{q \times 1}, (\alpha^{-1} - 1)\bSigma)$ captures micro-scale variation. The ``proportionality'' assumption for the variance of $\beps(\bs)$ will allow us to derive analytic posterior distributions using conjugate priors. 

The latent process $\bomega(\bs)$ captures the underlying spatial pattern and holds specific interest in many applications. Let $\bomega = \bomega(\calS) = [\bomega(\bs_1): \cdots : \bomega(\bs_n)]^\top$ 
{be the latent process on $\calS$, the parameter set of the latent model \eqref{eq: conj_latent_model} then becomes $\{\bbeta, \bomega, \bSigma\}$}. Letting $\bgamma^{\top} = [\bbeta^{\top}, \bomega^{\top}]$ be {$q\times (p+n)$}, we assume that $\{\bgamma, \bSigma\}\sim \mbox{MNIW}(\bmu_{\bgamma}, \bV_{\bgamma}, \bPsi, \nu)$, where $\bmu_{\bgamma}^\top = [\bmu_{\bbeta}^\top, \mathbf{0}_{q \times n}]$ and $\bV_{\bgamma} =  
\mbox{blockdiag}\{\bV_r, \brho_{\psi}(\calS, \calS)\}$. The posterior density is
\begin{equation}\label{eq: augmented_conj_post_v1}
\begin{aligned}
    p(\bgamma, \bSigma \given \bY) &\propto \mbox{MNIW}(\bgamma, \bSigma\given\bmu_{\bgamma}, \bV_{\bgamma}, \bPsi, \nu)\times \mbox{MN}_{n,q}(\bY_{n \times q} \given {[\bX : \bI_n]}\bgamma, {(\alpha^{-1} - 1)}\bI_n, \bSigma) \\
    &\propto \mbox{MNIW}(\bgamma, \bSigma\given \bmu_{\bgamma}^\ast, \bV^\ast, \bPsi^\ast, \nu^\ast)\;,
\end{aligned}
\end{equation}
where
\begin{equation}\label{eq: augmented_conj_post_v1_pars}
\begin{aligned}
    \bV^{\ast} &= \left[\begin{array}{cc} \frac{\alpha}{1 - \alpha} \bX^\top \bX + \bV_r^{-1} & \frac{\alpha}{1 - \alpha}\bX^\top \\ \frac{\alpha}{1 - \alpha}\bX & \brho_{\psi}^{-1}(\calS, \calS) + \frac{\alpha}{1 - \alpha}\bI_n \end{array} \right]^{-1}\;,\quad \bmu_{\bgamma}^\ast = \bV^{\ast} \left[\begin{array}{c} \frac{\alpha}{1 - \alpha}\bX^\top \bY + \bV_r^{-1}\bmu_{\bbeta} \\ \frac{\alpha}{1 - \alpha} \bY \end{array} \right],\\
    \bPsi^\ast &= \bPsi + \frac{\alpha}{1 - \alpha} \bY^\top\bY + \bmu_{\bbeta}^\top \bV_r^{-1}\bmu_{\bbeta} - \bmu_{\bgamma}^{\ast\top}\bV^{\ast-1} \bmu_{\bgamma}^\ast \; \mbox{  and  } \nu^\ast = \nu + n\; .
\end{aligned}
\end{equation}
For prediction on a set of locations $\calU$, we can estimate the unobserved latent process {$\bomega_\calU = \bomega(\calU) = [\bomega(\bu_1): \cdots: \bomega(\bu_{n'})]^\top$} and the response $\bY_\calU$ through 
\begin{multline}\label{eq: posterior_predict_augmented}
    p(\bY_\calU, \bomega_\calU \given \bY) = \int \mbox{MN}_{n',q}(\bY_\calU\given \bX_\calU \bbeta + \bomega_\calU,\; (\alpha^{-1} -1)\bI_{n'},\; \bSigma) \times \mbox{MN}_{n',q}(\bomega_\calU\given \bM_{\calU}\bomega, \bV_{\omega_{\calU}},\bSigma)\\  
    \times \mbox{MNIW}(\bgamma, \bSigma\given \bmu_{\bgamma}^\ast, \bV^\ast, \bPsi^\ast, \nu^\ast)\; d\bgamma d\bSigma\; ,
\end{multline}
where $\bM_{\calU} = \brho_{\psi}(\calU, \calS)\brho_{\psi}^{-1}(\calS,\calS)$ and $\bV_{\omega_{\calU}} = \brho_{\psi}(\calU, \calU) - \brho_{\psi}(\calU, \calS)\brho_{\psi}^{-1}(\calS,\calS)\brho_{\psi}(\calS,\calU)$. Posterior predictive inference proceeds by sampling one draw of $\bomega_\calU\sim \mbox{MN}_{n',q}(\bomega_\calU\given \bM_{\calU}\bomega, \bV_{\omega_{\calU}},\bSigma)$ for each posterior draw of $\{\bgamma,\bSigma\}$ and then one draw of $\bY_\calU \sim \mbox{MN}(\bX_\calU \bbeta + \bomega_\calU,(\alpha^{-1} -1)\bI_{n'},\; \bSigma)$ for each drawn $\{\bomega_\calU,\bgamma,\bSigma\}$. 

\subsection{Scalable Conjugate Bayesian Multivariate Models}\label{subsec: scalable_conj_models}
\paragraph{Conjugate multivariate response NNGP model} A conjugate Bayesian modeling framework is appealing for massive spatial data sets because the posterior distribution of the parameters are available in closed form circumventing the need for MCMC algorithms. The key computational bottleneck for Bayesian estimation of spatial process models concerns the computation and storage involving $\bkappa^{-1}$ in \eqref{eq: collapsed_spatial_pars1}. The required matrix computations require $\mathcal{O}(n^3)$ flops and $\mathcal{O}(n^2)$ storage
when $\bkappa$ is $n \times n$ and dense. While conjugate models reduce computational expenses by enabling direct sampling from closed-form posterior and posterior predictive distributions, the computation and storage of $\bkappa$ is still substantial for massive datasets.

One approach to circumvent the overwhelming computations is to develop a sparse alternative for $\bkappa^{-1}$ in \eqref{eq: collapsed_spatial_pars1}. One such approximation that has generated substantial recent attention in the spatial literature is an approximation due to \cite{ve88}. Consider the spatial covariance matrix $\bkappa = \brho_{\psi} + \delta_{\bs=\bs'}\bI_n$ in (\ref{eq: MN_density}). This is a dense $n\times n$ matrix with apparently no exploitable structure. Instead, we specify a sparse Cholesky representation \begin{equation}\label{eq: NNGP_approx_invK}
    \bkappa^{-1} = (\bI - \bA_{\bkappa})^\top \bD_{\bkappa}^{-1} (\bI - \bA_{\bkappa})\; ,
\end{equation}
where $\bD_{\bkappa}$ is a diagonal matrix and $\bA_{\bkappa}$ is a sparse lower-triangular matrix with $0$ along the diagonal and with no more than a fixed small number $m$ of nonzero entries in each row of $\bA_{\bkappa}$. The diagonal entries of $\bD_{\bkappa}$ and the nonzero entries of $\bA_{\bkappa}$ are obtained from the conditional variance and conditional expectations for a Gaussian process with covariance function $\rho_{\psi}(\bs,\bs')$. 

{Vecchia's approximation begins with a fixed ordering of the locations in $\calS$ and approximates the joint density of $\bY$ as
\begin{equation}\label{eq: vecchia_approx}
p(\bY) = p(\by(\bs_1))\prod_{i=2}^n p(\by(\bs_i)\given \by(H(\bs_i))) \approx p(\by(\bs_1))\prod_{i=2}^n p(\by(\bs_i)\given \by(N_m(\bs_i)))\;,
\end{equation}
where $H(\bs_i) = \{\bs_1,\bs_2,\ldots,\bs_{i-1}\}$, $N_m(\bs_i) = H(\bs_i)$ for $i=2,\ldots,m$ and $N_m(\bs_i)\subset H(\bs_i)$ for $i=m+1,\ldots,n$ comprises at most $m$ neighbors of of $\bs_i$ among locations $\bs_j\in \calS$ such that $j<i$. Also, $\by(\calA)$ for any set $\calA \subseteq \calD$ is the collection of $\by(\bs_i)$'s for $\bs_i\in A$. Practical advice has been provided in \cite{stein04} and simple ordering by either the $x$-coordinate or the $y$-coordinate or the sum $x+y$ are all practicable solutions for massive data. A more formal algorithm based upon theoretical insights on the effect of ordering on inference has been provided by \cite{guinness2018permutation}. Here we focus on constructing (\ref{eq: NNGP_approx_invK}) using a fixed ordering of the locations in $\mathcal{S}$.} 

The matrix $\bkappa^{-1}$ in (\ref{eq: NNGP_approx_invK}) can be derived from (\ref{eq: vecchia_approx}) using standard results in multivariate Gaussian distribution theory \citep[see, e.g.][]{banerjee2017high}. The $(i,j)$-th entry of $\bA_{\bkappa}$ is $0$ whenever $\bs_j \notin N_m(\bs_i)$. This means that each row of $\bA_{\kappa}$ contains at most $m$ nonzero entries. Suppose $i_1 < i_2 < \ldots < i_m$ are the $m$ column indices that contain nonzero entries in the $i$-th row of $\bA_{\bkappa}$. 
{Let $\bA_{\bkappa} = [\ba_1:\cdots:\ba_n]^\top$ and $\bD_{\bkappa} = \mbox{diag}(d_1, d_2,\ldots, d_n)$, where $d_1 = \alpha^{-1}$. The first row of $\bA$ has all elements equal to $0$ and for $i = 2, \ldots, n$ we obtain
\begin{equation}\label{eq: NNGP_a_d_construct}
\begin{split}
    [\{\ba_i\}_{i_1}, \ldots, \{\ba_i\}_{i_m}] &= \brho_{\psi}(\bu_i, N_m(\bs_i))\{\brho_{\psi}(N_m(\bs_i), N_m(\bs_i)) + (\alpha^{-1} - 1) \bI_{m}\}^{-1}\; ,\\
   d_i &= \alpha^{-1} - [\{\ba_i\}_{i_1}, \ldots, \{\ba_i\}_{i_m}]\brho_{\psi}(N_m(\bs_i), \bu_i)\; ,
\end{split}
\end{equation}
where $\{\cdot\}_{k}$ denotes the $k$-th element of a vector.}
Equation~(\ref{eq: NNGP_a_d_construct}) completely specifies $\bA_{\bkappa}$ and $\bD_{\kappa}$ and, hence, a sparse $\bkappa^{-1}$ in (\ref{eq: NNGP_approx_invK}). The construction in (\ref{eq: NNGP_a_d_construct}) can be implemented in parallel and requires storage or computation of matrices of sizes no greater than $m\times m$, where $m << n$, and costs $\mathcal{O}(n)$ flops and storage. 

Based on Section~\ref{sec: conj_resp}, the posterior distribution $\bbeta, \bSigma \given \bY$ follows $\mbox{MNIW}(\bmu^\ast, \bV^\ast, \bPsi^\ast, \nu^\ast)$ where $\{\bmu^\ast, \bV^\ast, \bPsi^\ast, \nu^\ast\}$ are given in \eqref{eq: collapsed_spatial_pars1}. With the sparse representation of $\bkappa^{-1}$ in \eqref{eq: NNGP_approx_invK}, the process of obtaining posterior inference for $\{\bbeta, \bSigma\}$ only involves steps with storage and computational requirement in $\mathcal{O}(n)$. 

{To obtain the predictions at arbitrary (unobserved) locations $\calU = \{\bu_1, \ldots, \bu_{n'}\}$, we follow \cite{datta16} and extend the approximation in (\ref{eq: vecchia_approx}) to the Nearest Neighbor Gaussian Process (NNGP). We} extend the definition of $N_m(\bs_i)$'s to arbitrary locations in $\calU$ by defining $N_m(\bu_i)$ to be the set of $m$ nearest neighbors of $\bu_i$ from $\calS$. Furthermore, we assume that $\by(\bu)$ and $\by(\bu')$ are conditionally independent of each other given {$\bY = \by(\mathcal{S})$} and the other model parameters. Thus, for any $\bu_i\in \calU$, we have
\begin{equation}
    \by(\bu_i)\given \bY, \bbeta, \bSigma \sim \mbox{N}(\bbeta^{\top}\bx(\bu_i) + {[\bY - \bX\bbeta]^\top\Tilde{\ba}_i},
    \, \Tilde{d}_i  \bSigma), \; i = 1, \ldots, n'\; ,
\end{equation}
{where $\Tilde{\ba}_i$ is an $n \times 1$ vector with $m$ non-zero elements. If 
$N_m(\bu_i) = \{\bs_{i_k}\}_{k = 1}^m$, then} 
\begin{equation}\label{eq: NNGP_predict_collapsed_par}
\begin{split}
(\{\Tilde{\ba}_i\}_{i_1}, &\ldots, \{\Tilde{\ba}_i\}_{i_m}) = \brho_{\psi}(\bu_i, N_m(\bu_i))\{\brho_{\psi}(N_m(\bu_i), N_m(\bu_i)) + (\alpha^{-1} - 1) \bI_{m}\}^{-1}\; ,\\
\Tilde{d}_i &= \alpha^{-1} - \brho_{\psi}(\bu_i, N_m(\bu_i))[\brho_{\psi}(N_m(\bu_i), N_m(\bu_i)) + (\alpha^{-1} - 1) \bI_{m}]^{-1}\brho_{\psi}(N_m(\bu_i), \bu_i)\; .
\end{split}
\end{equation}
If $\Tilde{\bA} = [\Tilde{\ba}_1: \cdots : \Tilde{\ba}_{n'}]^\top$ and $\Tilde{\bD} = \mbox{diag}(\{\Tilde{d}_i\}_{i = 1}^n)$, then the conditional predictive density for $\bY_\calU$ is 
\begin{equation}\label{eq: resp_NNGP_YU}
 \bY_\calU \given \bY, \bbeta, \bSigma \sim \mbox{MN}(\bX_\calU\bbeta + \Tilde{\bA}[\bY - \bX\bbeta], \Tilde{\bD}, \bSigma)  \; . 
\end{equation} 
Since the posterior distribution of $\{\bbeta,\bSigma\}$ and the conditional predictive distribution of $\bY_\calU$ are both available in closed form, direct sampling from the posterior predictive distribution is straightforward. A detailed algorithm for obtaining the posterior inference on parameter set $\{\bbeta, \bSigma\}$ and the posterior prediction over a new set of locations $\calU$ is given as below. 

\noindent{ \rule{\textwidth}{1pt}
{\fontsize{8}{8}\selectfont
	\textbf{Algorithm~1}: 
	Obtaining posterior inference of $\{\bbeta, \bSigma\}$ and predictions on $\calU$ for conjugate multivariate response NNGP model \\
	\rule{\textwidth}{1pt}\\
	\begin{enumerate}
		\item Construct $\bV^\ast$, $\bmu^\ast$, $\bPsi^\ast$ and $\nu^\ast$: 
		\begin{enumerate}
		    \item Compute $\bL_r$ the Cholesky decomposition of $\bV_r$ \hfill{$\mathcal{O}(p^3)$ flops}
		    \item Compute $\bD\bI\bA\bX = \bD_{\bkappa}^{-\frac{1}{2}}(\bI -\bA_{\bkappa}) \bX$ and $\bD\bI\bA\bY = \bD_{\bkappa}^{-\frac{1}{2}}(\bI -\bA_{\bkappa})\bY$ 
		    \begin{itemize}
			    \item  
			    {Construct $\bA_{\bkappa}$ and $\bD_{\bkappa}$ as described in \eqref{eq: NNGP_a_d_construct} \hfill{$\mathcal{O}(nm^3)$ flops} 
			    }
			    \item Compute $\bD\bI\bA\bX = \bD_{\bkappa}^{-\frac{1}{2}}(\bI -\bA_{\bkappa})\bX$ and $\bD\bI\bA\bY = \bD_{\bkappa}^{-\frac{1}{2}}(\bI -\bA_{\bkappa})\bY$ by $\bA_{\bkappa}$ and $\bD_{\bkappa}$ \hfill{$\mathcal{O}(nm(p + q))$ flops}
		    \end{itemize}
		    \item Obtain $\bV^\ast$, $\bmu^\ast$ and $\bPsi^\ast$
		    \begin{itemize}
		        \item Compute $\bV^\ast = (\bD\bI\bA\bX^\top\bD\bI\bA\bX + \bV_r^{-1})^{-1}$ and its Cholesky decomposition $\bL_{v\ast}$ \hfill{$\mathcal{O}(np^2)$ flops}
		        \item Compute $\bmu^\ast = \bV^{\ast}(\bD\bI\bA\bX^\top\bD\bI\bA\bY + \bV_r^{-1}\bmu_{\bbeta})$
		        \hfill{$\mathcal{O}(npq)$ flops}
		        \item Compute $\bPsi^\ast =\bPsi + \bD\bI\bA\bY^\top\bD\bI\bA\bY + (\bL_r^{-1} \bmu_{\bbeta})^\top (\bL_r^{-1} \bmu_{\bbeta}) - (\bL_{v\ast}^{-1} \bmu^\ast)^\top (\bL_{v\ast}^{-1} \bmu^\ast)$
		        \hfill{$\mathcal{O}(nq^2)$ flops}
		        \item Compute $\nu^\ast = \nu + n$ \hfill{$\mathcal{O}(1)$ flops}
		    \end{itemize}
		\end{enumerate}
		\item Generate posterior samples $\{\bY_\calU^{(l)}\}_{l = 1}^L$ on a new set $\calU$ given $\bX_\calU$ 
		\begin{enumerate}
		\item Construct $\Tilde{\bA}$ and $\Tilde{\bD}$ as described in \eqref{eq: NNGP_predict_collapsed_par} \hfill{$\mathcal{O}(n'm^3)$ flops}
		\item \text{For} $l$ in $1:L$
			\begin{enumerate}
		    \item Sample $\bSigma^{(l)} \sim \mathrm{IW}(\bPsi^\ast, \nu^\ast)$   
		    \hfill{$\mathcal{O}(q^3)$ flops}
		    \item Sample $\bbeta^{(l)} \sim \mbox{MN}(\bmu^\ast, \bV^\ast, \bSigma^{(l)})$
		    \begin{itemize}
			\item Calculate Cholesky decomposition of $\bSigma^{(l)}$,  $\bSigma^{(l)} = \bL_{\Sigma^{(l)}}\bL^\top_{\Sigma^{(l)}}$ \hfill{$\mathcal{O}(q^3)$ flops} 
			\item Sample $\bu \sim \mbox{MN}(\mathbf{0}_{p \times q}, \bI_p, \bI_q)$ (i.e. $\mbox{vec}(\bu)\sim \mbox{MVN}(\mathbf{0}_{pq \times 1}, \bI_{pq})$)
			\hfill{$\mathcal{O}(pq)$ flops}
			\item Generate $\bbeta^{(l)} = \bmu^\ast + \bL_{v\ast}\bu \bL_{\bSigma^{(l)}}^\top$ \hfill{$\mathcal{O}(p^2q + pq^2)$ flops} 
		    \end{itemize}
		    \item Sample $\bY_\calU^{(l)} \sim \mbox{MN}(\bX_\calU\bbeta^{(l)} + \Tilde{\bA}[\bY - \bX\bbeta^{(l)}], \Tilde{\bD}, \bSigma^{(l)})$
		    \begin{itemize}
		        \item Sample $\bu \sim \mbox{MN}(\mathbf{0}_{n' \times q}, \bI_{n'}, \bI_q)$. \hfill{$\mathcal{O}(n'q)$ flops}
		        \item Generate $\bY_\calU^{(l)} = \bX_\calU\bbeta^{(l)} + \Tilde{\bA}[\bY - \bX\bbeta^{(l)}] + \Tilde{\bD}^{\frac{1}{2}}\bu\bL_{\bSigma^{(l)}}^\top$ \hfill{$\mathcal{O}((n'+n)pq + n'(q^2 + mq))$ flops 
		        }
		    \end{itemize}
		\end{enumerate}
		\end{enumerate}
	\end{enumerate}	
	\vspace*{-8pt}
	\rule{\textwidth}{1pt}
} }

\paragraph{Conjugate multivariate latent NNGP model} Bayesian estimation for the conjugate multivariate latent model is more challenging because inference is usually sought on the (high-dimensional) latent process itself. In particular, the calculations involved in $\bV^\ast$ in \eqref{eq: augmented_conj_post_v1} are often too expensive for large data sets even when the precision matrix $\brho^{-1}_\psi(\calS, \calS)$ is sparse. Here, the latent process $\bomega(\bs)$ in (\ref{eq: conj_latent_model}) follows a multivariate Gaussian process so that its realizations over $\calS$ follows $\bomega \sim \mbox{MN}(\mathbf{0}_{n \times q}, \Tilde{\brho}, \bSigma)$, where $\Tilde{\brho}$ is the Vecchia approximation of $\brho_{\psi}(\calS, \calS)$. Hence, $\Tilde{\brho}^{-1} = (\bI - \bA_{\brho})^{\top}\bD_{\brho}^{-1}(\bI - \bA_{\brho})$, where $\bA_{\brho}$ and $\bD_{\brho}$ are constructed analogous to $\bA_{\bkappa}$ and $\bD_{\bkappa}$ in (\ref{eq: NNGP_approx_invK}) with 
{$\alpha$ replaced by $1$}. This corresponds to modeling $\bomega(\bs)$ as an NNGP \citep[see, e.g.,][for details on the NNGP and its properties]{datta16, datta16b, banerjee2017high}. 
{The distribution theory for $\bomega(\bs)$ over $\calS$ and $\calU$ follows analogous to that of $\by(\bs)$ in the previous section.} 

The posterior distribution of $\{\bgamma, \bSigma\}$ follows a Matrix-Normal distribution similar to (\ref{eq: augmented_conj_post_v1}), but with $\brho_{\psi}(\calS,\calS)^{-1}$ in (\ref{eq: augmented_conj_post_v1_pars}) replaced by its Vecchia approximation $\Tilde{\brho}_{\psi}(\calS,\calS)$. 
However, sampling $\{\bgamma, \bSigma\}$ is still challenging for massive data sets, where we seek to minimize storage and operations with large matrices. Here we introduce a useful representation. Let $\bV_{\brho}$ be a non-singular square matrix such that $\brho_{\psi}^{-1}(\calS, \calS) = \bV_{\brho}^\top\bV_{\brho}$ where we write $\bV_{\brho} = \bD_{\brho}^{-1/2}(\bI-\bA_{\brho})$. We treat the prior of $\bgamma$ as additional ``observations'' and recast $p(\bY, \bgamma \given \bSigma) = p(\bY \given \bgamma, \bSigma) \times p(\bgamma \given \bSigma)$ into an augmented linear model
\begin{equation}\label{eq: augment_linear_latent}
\begin{array}{c}
\underbrace{ \left[ \begin{array}{c} \sqrt{\frac{\alpha}{1 - \alpha}} \bY\\ \bL_r^{-1} \bmu_{\bbeta} \\ \mathbf{0}_{n \times q} \end{array} \right]}_{\bY^{*}}
= \underbrace{ \left[ \begin{array}{cc} \sqrt{\frac{\alpha}{1 - \alpha}} \bX & \sqrt{\frac{\alpha}{1 - \alpha}} \bI_n \\ \bL_r^{-1}& \mathbf{0}_{p \times n} \\  \mathbf{0}_{n \times p}& \bV_{\brho} \end{array} \right] }_{\bX^{*}}
\underbrace{ \left[ \begin{array}{c} \bbeta \\ \bomega \end{array} \right]}_{\bgamma}+ \underbrace{ \left[ \begin{array}{c} \etab_1 \\ \etab_2 \\ \etab_3 \end{array} \right]}_{\etab}
\end{array} ,
\end{equation}
where $\bL_r$ is the Cholesky decomposition of $\bV_r$, and $\etab \sim \mbox{MN}(\mathbf{0}_{(2n+p) \times q}, \bI_{2n + p}, \bSigma)$. With a flat prior for $\bbeta$, $\bL_r^{-1}$ degenerates to $\bO$ and does not contribute to the linear system. Equation~\eqref{eq: augmented_conj_post_v1_pars} simplifies to
\begin{equation}\label{eq: augmented_conj_post_v2_pars}
\begin{aligned}
\bV^\ast &= (\bX^{\ast\top}\bX^\ast)^{-1}\; , \;
\bmu^\ast = (\bX^{\ast\top}\bX^\ast)^{-1}\bX^{\ast\top}\bY^\ast \; ,\\
\bPsi^\ast &= \bPsi + (\bY^\ast - \bX^\ast \bmu^\ast)^\top(\bY^\ast - \bX^\ast \bmu^\ast)\; , \;
\nu^\ast = \nu + n \; .
\end{aligned}
\end{equation}
Following developments in \citet{zdb2019} for the univariate case, one can efficiently generate posterior samples through a conjugate gradient algorithm exploiting the sparsity of $\bV_{\brho}$. The sampling process for $\bgamma$ will be scalable when there is a sparse precision matrix $\brho_\psi^{-1}(\calS, \calS)$. It is also possible to construct $\bV^\ast$ and $\bmu^\ast$ in \eqref{eq: augmented_conj_post_v2_pars} using $\brho_\psi^{-1}(\calS, \calS)$ instead of $\bV_{\brho}$. We refer to \citet{zdb2019} for further details of this construction. We provide a detailed algorithm of the conjugate multivariate latent NNGP model in 
{Algorithm~2. We solve the linear system $\bX^{\ast\top}\bX^\ast\bmu^\ast = \bX^{\ast\top}\bY^\ast$ for $\bmu^\ast$, compute $\{\bPsi^\ast, \nu^\ast\}$ and generate posterior samples of $\bSigma$ from $\mbox{IW}(\bPsi^\ast, \nu^\ast)$. Posterior samples of $\bgamma$ are obtained by generating $\etab \sim \mbox{MN}(\mathbf{0}_{ (2n+p) \times q}, \bI_{2n + p}, \bSigma)$, solving $\bX^{\ast\top}\bX^\ast \bv = \bX^{\ast\top}\etab$ for $\bv$ and then obtaining posterior samples of $\bgamma$ from $\bgamma = \bmu^\ast + \bv$. }
We implement a ``Sparse Equations and Least Squares'' (LSMR) algorithm \citep{fong2011lsmr} to solve the linear system $\bX^{\ast\top}\bX^\ast\bmu^\ast = \bX^{\ast\top}\bY^\ast$ and $\bX^{\ast\top}\bX^\ast \bv = \bX^{\ast\top}\etab$ needed to generate $\bgamma$. LSMR is a conjugate-gradient type algorithm for solving sparse linear equations $\bA \bx = \mathbf{b}$ where the matrix $\bA$ may be square or rectangular. The matrix $\bA := \bX^\ast$ is a sparse tall matrix. LSMR only requires storing  $\bX^{\ast}$, $\bY^{\ast}$ and $\etab^{\ast}$ and, unlike the conjugate gradient algorithm, avoids $\bX^{\ast\top}\bX^{\ast}$, $\bX^{\ast\top}\bY$ and $\bX^{\ast\top}\etab$. LSMR also tends to produce more stable estimates than conjugate gradient. We have also tested a variety of conjugate gradient methods and preconditioning methods, where we have observed that their performances varied across different data sets. The LSMR without conditioning showed a relatively good performance for the latent models. 
Therefore, we choose LSMR without preconditioning for our current illustrations.
Posterior predictive inference will adapt from (\ref{eq: posterior_predict_augmented}) for scalable models. After sampling $\{\bgamma,\bSigma\}$, we sample one draw of $\bomega_\calU\sim \bomega_\calU\given \bgamma, \bSigma \sim \mbox{MN}([\mathbf{0}_{n' \times p}, \Tilde{\bA}]\bgamma, \Tilde{\bD}, \bSigma)$ for each sampled $\{\bgamma,\bSigma\}$, where $\Tilde{\bA} = [\Tilde{\ba}_1: \cdots: \Tilde{\ba}_n]^\top$, $\Tilde{\bD} = \mbox{diag}(\{\Tilde{d}_i\}_{i = 1}^n)$ with 
\begin{equation}\label{eq: conj_latent_NNGP_A_D_U}
	\begin{split}
   \left[\{\Tilde{\ba}_i\}_{i_1}, \ldots, \{\Tilde{\ba}_i\}_{i_m}\right] &=  \brho_{\psi}(\bu_i, N_m(\bu_i))\brho_{\psi}^{-1}(N_m(\bu_i), N_m(\bu_i))\;,\\
    \Tilde{d}_i &= 1 - \brho_{\psi}(\bu_i, N_m(\bu_i))\brho_{\psi}^{-1}(N_m(\bu_i), N_m(\bu_i))\brho_{\psi}(N_m(\bu_i),\bu_i)\;.
    \end{split}
\end{equation}
Finally, for each sampled $\{\bbeta,\bomega_\calU,\bSigma\}$ we make one draw of $\bY_\calU\sim \mbox{MN}(\bX_\calU\bbeta+\bomega_\calU, {(\alpha^{-1} - 1)}\bI_{n'}, \bSigma)$.
The following algorithm provides the steps for predictive inference.

\noindent{ \rule{\textwidth}{1pt}
{\fontsize{8}{8}\selectfont
	\textbf{Algorithm~2}: 
	Obtaining posterior inference of $\{\bgamma, \bSigma\}$ and predictions on set $\calU$ for conjugate multivariate latent NNGP\\
	\rule{\textwidth}{1pt}
	\begin{enumerate}
		\item Construct $\bX^\ast$ and $\bY^\ast$ in \eqref{eq: augment_linear_latent}
		\begin{enumerate}
		    \item $\bL_r^{-1}$ and $\bL_r^{-1} \boldsymbol{\mu_\beta}$
		    \begin{itemize}
			    \item Compute the Cholesky decomposition of $\bV_r$, $\bL_r$ \hfill{$\mathcal{O}(p^3)$ flops}
			    \item Compute $\bL_r^{-1}$ and $\bL_r^{-1} \bmu_{\bbeta}$    \hfill{$\mathcal{O}(p^2q)$ flops}
		    \end{itemize}
		    \item $\bV_{\brho}$ 
		    \begin{itemize}
			    \item  Construct $\bA_{\brho}$ and $\bD_{\brho}$ 
			    \hfill{$\mathcal{O}(nm^3)$ flops}
			    \item Compute $\bV_{\brho} = \bD_{\brho}^{-\frac{1}{2}}(\bI - \bA_{\brho})$  \hfill{$\mathcal{O}(nm)$ flops}
		    \end{itemize}
		    \item Construct $\bX^\ast$ and $\bY^\ast$  
		\end{enumerate}
		\item Obtain $\bmu^\ast$,   $\bPsi^\ast$ and $\nu^\ast$. 
		\begin{enumerate}
		    \item Obtain $\bmu^\ast = [\bmu^\ast_1: \cdots: \bmu^\ast_q]$
		    \begin{itemize}
		        \item Solve $\bmu^{\ast}_i$ from $\bX^{\ast}\bmu^{\ast}_i = \bY^{\ast}_i$ by LSMR for $i = 1, \ldots, q$. 
		    \end{itemize}
		    \item Obtain $\bPsi^\ast$ and $\nu^\ast$
		    \begin{itemize}
		        \item Generate $\bu = \bY^\ast - \bX^\ast\bmu^\ast$ \hfill{$\mathcal{O}(nq(p + m))$ flops}
		        \item Compute $\bPsi^\ast = \bPsi + \bu^\top\bu$\hfill{$\mathcal{O}(nq^2)$ flops 
		        }
		        \item Compute $\nu^\ast = \nu + n$ \hfill{$\mathcal{O}(1)$ flops}
		    \end{itemize}
		\end{enumerate}
		\item Generate posterior samples of $\{\bgamma^{(l)}, \bSigma^{(l)}\}_{l = 1}^L$. For $l$ in $1:L$
		\begin{enumerate}
		    \item Sample $\bSigma^{(l)} \sim \mathrm{IW}(\bPsi^\ast, \nu^\ast)$  \hfill{$\mathcal{O}(q^3)$ flops}
		    \item Sample $\bgamma^{(l)} \sim \mbox{MN}(\bmu^\ast, \bV^\ast, \bSigma^{(l)})$
		    \begin{itemize}
			\item Sample $\bu \sim \mbox{MN}(\mathbf{0}_{(2n+p) \times q}, \bI_{2n + p}, \bI_q)$ \hfill{$\mathcal{O}(nq)$ flops}
			\item Calculate Cholesky decomposition of $\bSigma^{(l)}$, $\bSigma^{(l)} = \bL_{\bSigma^{(l)}}\bL^\top_{\bSigma^{(l)}}$  
			\hfill{$\mathcal{O}(q^3)$ flops}
			\item Generate $\etab = \bu\bL^{(l)\top} = [\etab_1 : \cdots :\etab_q]$
			\hfill{$\mathcal{O}(nq^2)$ flops}
			\item Solve $\bv_i$ from $\bX^\ast\bv_i = \etab_i$ by LSMR for $i = 1, \ldots, q$.
			\item Generate $\bgamma^{(l)} = \bmu^\ast + \bv$ with $\bv = [\bv_1: \cdots:\bv_q]$ 
			\hfill{$\mathcal{O}(nq)$ flops 
			}
		    \end{itemize}
		\end{enumerate}
		\item Generate posterior samples of $\{\bY_\calU^{(l)}\}$ on a new set $\calU$ given $\bX_\calU$.
		\begin{enumerate}
		\item Construct $\Tilde{\bA}$ and $\Tilde{\bD}$ using \eqref{eq: conj_latent_NNGP_A_D_U} \hfill{$\mathcal{O}(n'm^3)$ flops}
		\item For $l$ in $1:L$
		\begin{enumerate}
		    \item Sample $\bomega_\calU^{(l)} \sim \mbox{MN}([\mathbf{0}_{n'\times p}, \Tilde{\bA}]\bgamma^{(l)}, \Tilde{\bD}, \bSigma^{(l)})$
		\begin{itemize}
		\item Sample $\bu \sim \mbox{MN}(\mathbf{0}_{n' \times q}, \bI_{n'}, \bI_q)$ \hfill{$\mathcal{O}(n'q)$ flops}
		\item Generate $\bomega_\calU^{(l)} = [\mathbf{0}_{n'\times p}, \Tilde{\bA}]\bgamma^{(l)} + \Tilde{\bD}^{\frac{1}{2}}\bu\bL_{\bSigma^{(l)}}^\top$ \hfill{$\mathcal{O}(n'mq+ n'q^2)$ flops  
		}
		\end{itemize} 
		\item Sample $\bY_\calU^{(l)} \given \bomega_\calU^{(l)}, \bgamma^{(l)}, \bSigma^{(l)} \sim \mbox{MN}(\bX_\calU\bbeta+\bomega_\calU, (\alpha^{-1} - 1)\bI_{n'}, \bSigma)$ 
		\begin{itemize}
		\item Sample $\bu \sim \mbox{MN}(\mathbf{0}_{n' \times q}, \bI_{n'}, \bI_q)$  \hfill{$\mathcal{O}(n'q)$ flops}
		\item Generate $\bY_\calU^{(l)} = \bX_\calU\bbeta + \bomega_\calU^{(l)} + {(\alpha^{-1} - 1)} \bu \bL_{\bSigma^{(l)}}^\top$ \hfill{$\mathcal{O}(n'pq + n'q^2)$ flops 
		}
		\end{itemize}
		\end{enumerate}
		\end{enumerate}
	\end{enumerate}	
	\vspace*{-8pt}
	\rule{\textwidth}{1pt}
} }

{
\paragraph{Model comparisons} We will use posterior predictive performance as a key measure to compare inferential performance among the multivariate spatial models. For multivariate models we investigate model fit for each variable as well as by combining across variables. For example, using a common hold-out set $\{\bu_1,\bu_2,\ldots,\bu_{n'}\}$ for each model we evaluate the root mean squared prediction error for each outcome as $\mbox{RMSPE} = \sqrt{\sum_{i = 1}^{n'}(y_j(\bu_i) - \hat{y_j}(\bu_i))^2/{n'}}$ for $j = 1,\ldots, q$ and also for all the responses combined as $\mbox{RMSPE} = \sqrt{\sum_{j=1}^q\sum_{i = 1}^{n'}(y_j(\bu_i) - \hat{y_j}(\bu_i))^2/(n'q)}$, where $y_j(\bu_i)$ and $\hat{y}_j(\bu_i)$ are the observed and predicted (posterior predictive mean) values of the outcome, respectively. Other metrics we compute are the prediction interval coverage (CVG; the percent of intervals containing the true value), interval coverage for intercept-centered latent processes of observed response (CVGL), and the mean continuous rank probability score ($\mbox{MCRPS} = \sum_{i = 1}^{n'} \mbox{CRPS}_j(\bu_i)/{n'}$ for each $j= 1, \ldots, q$, where $\mbox{CPRS}_j(\bu_i)$ is the CRPS of the $j$-th response on held location $\bu_i$ \citep{gneiting2007strictly}). 
To calculate $\mbox{CRPS}_j(\bu_i)$, we approximate the predictive distribution by a Normal distribution with mean centered at the predicted value {$\hat{y}_j(\bu_i)$} and standard deviation equal to the predictive standard error $\hat{\sigma}_j(\bu_i)$. Therefore, $\mbox{CPRS}_j(\bu_i) = \hat{\sigma}_j(\bu_i)[1/\sqrt{\pi} - 2\varphi(z_{ij}) - z_{ij} (2 \Phi(z_{ij})-1)]$ where $z_{ij} = (y_j(\bu_i) - \hat{y}_j(\bu_i) ) / \hat{\sigma}_j(\bu_i)$, and $\varphi$ and $\Phi$ are the density and the cumulative distribution function of a standard Normal variable, respectively.} 

{In simulation experiments, where we know the values of true parameters (and the latent process) generating the data, we also evaluate the mean squared error of intercept-centered latent processes $\mbox{MSEL}= {\sum_{i = 1}^n(\omega_j(\bs_i) + \bx(\bs_i)^{\top}\bbeta_{j}- \hat{\omega_j}(\bs_i) - \hat{\bbeta}_{j})^2/n}$ for each $j = 1, \ldots, q$ and for all responses combined as $\mbox{MSEL}= {\sum_{j=1}^q\sum_{i = 1}^n(\omega_j(\bs_i) + \bx(\bs_i)^{\top}\bbeta_{j}- \hat{\omega_j}(\bs_i) - \hat{\bbeta}_{j})^2/(nq)}$, where $\hat{\omega_j}(\bs_i)$ and $\hat{\bbeta}_{j}$ are the posterior mean estimates of the latent process and regression slopes, respectively, obtained by analyzing the simulated data. For NNGP models there is also the question of sensitivity to the number of neighbors $m$. Here, we rely upon the recommendations in \cite{datta16} based upon extensive simulation experiments that for standard covariance functions usually $m = 10$ nearest neighbors suffice to deliver robust substantive inference even for massive data sets. \cite{datta2016nearest} investigated NNGP models with an unknown $m$ that was modeled using a discrete prior, but reported almost indistinguishable inference from that obtained by fixing $m$. Hence, we do not pursue modeling $m$ here.}
 
\subsection{Cross-validation for Conjugate Multivariate NNGP Models}\label{subsec: CV_conj_NNGP}
Conjugate Bayesian multivariate regression models depends on fixed hyperparameters in the model. 
We adapt a univariate $K$-fold cross-validation algorithm for choosing $\{\psi, \alpha\}$ 
\citep{finley2019efficient} to our multivariate setting. For each $\{\psi,\alpha\}$ on a grid of candidate values we fit the conjugate model and perform predictive inference. We compare the model predictions and choose the $\{\psi,\alpha\}$ that produces the 
{best performance for some model fitting criterion, e.g.,} the least magnitude of root mean square prediction error (RMSPE). 
{Specifically, we divide the data into $K$ folds (holdout sets), say $\calS_1,\ldots,\calS_K$, and predict $\by(\bs)$ over each of these folds using the remaining data (outside of the fold) for training. We calculate the RMSPE over the $K$ folds as $\sum_{k=1}^K\sqrt{\sum_{s\in \calS_k}\|\by(\bs) - \hat{\by}(\bs)\|^2/(|\calS_k|q)}$, where $\hat{\by}(\bs)$ is the posterior predictive mean of $\by(\bs)$ and $|\calS_k|$ is the number of locations in $\calS_k$. This completes the cross-validation exercise for one choice of $\{\psi,\alpha\}$. We repeat this for all $\{\psi,\alpha\}$ over the grid and choose the point that corresponds to the smallest RMSPE.} Inference corresponding to this choice of hyperparameters is then presented. This is appealing for scalable Gaussian process models, which, for any fixed $\{\psi, \alpha\}$, can deliver posterior inference rapidly at new locations requiring storage and flops in $\mathcal{O}(n)$ only. The next algorithm presents the details of the steps for implementing $K$-fold cross-validation to choose the hyperparameters.  

\noindent{ \rule{\textwidth}{1pt}
{\fontsize{8}{8}\selectfont
	\textbf{Algorithm~3}: Cross-validation of tuning $\psi$, $\alpha$ for conjugate multivariate response or latent NNGP model\\
	\rule{\textwidth}{1pt}
	\begin{enumerate}
		\item[1.] Split $\calS$ into $K$ folds, and build the neighbor index. 
		\begin{itemize}
		    \item Split $\calS$ into $K$ folds $\{\calS_k\}_{k = 1}^K$, where $\calS_{-k}$ denotes the set of locations in $\calS$ that are not included in $\calS_k$.
			\item Build nearest neighbors for $\{\calS_{-k}\}_{k = 1}^K$
			\item Find the collection of nearest neighbor set for $\calS_k$ among $\calS_{-k}$ for $k = 1, \ldots, K$.
		\end{itemize} 
		\item[2.] (For response NNGP) Fix $\psi$ and $\alpha$, obtain posterior mean of $\bbeta$ after removing the $k^{th}$ fold of the data:
		\begin{itemize}
			\item Use step 1 in Algorithm~1 to obtain $\hat{\bbeta}_k$ by taking $\calS$ to be $\calS_{-k}$ and $\bmu^\ast$ to be $\hat{\bbeta}_k$.
		\end{itemize}
		\item[  ] (For latent NNGP) Fix $\psi$ and $\alpha$, obtain posterior mean of $\bgamma_k= \{ \bbeta, \bomega(\calS_{-k}) \}$ after removing the $k^{th}$ fold of the data:
		\begin{itemize}
			\item Use step 1-2 in Algorithm~3 to obtain $\hat{\bgamma}_k$ by taking $\calS$ to be $\calS_{-k}$ and $\bmu^\ast$ to be $\hat{\bgamma}_k$.
		\end{itemize}
		\item[3.] (For response NNGP) Predict posterior means of $\by(\calS_{k})$
		\begin{itemize}
			\item Construct matrix $\Tilde{\bA}$ through \eqref{eq: NNGP_predict_collapsed_par} by taking $\calS$ to be $\calS_{-k}$ and $\calU$ to be $\calS_{k}$.
			\item According to \eqref{eq: resp_NNGP_YU}, the predicted posterior mean of $\by(\calS_k)$ follows\\
			$\hat{\by}(\calS_k) = \mbox{E}[\by(\calS_k) \given \by(\calS_{-k})] = \bx(\calS_{k})\hat{\bbeta}_k + \Tilde{\bA}[\by(\calS_{-k}) - \bx(\calS_{-k})\hat{\bbeta}_k]$
		\end{itemize}
		\item[  ] (For latent NNGP) Predict posterior means of $\by(\calS_{k})$
		\begin{itemize}
			\item Construct matrix $\Tilde{\bA}$ by taking $\calS$ to be $\calS_{-k}$ and $\calU$ to be $\calS_{k}$.
			\item The predicted posterior mean of $\by(\calS_k)$ follows\\
			$\hat{\by}(\calS_k) = \mbox{E}[\by(\calS_k) \given \by(\calS_{-k})] = \mbox{E}_{\bomega}[\mbox{E}_{\by}[\by(\calS_{k}) \given \bomega(\calS_{-k}), \by(\calS_{-k})]] = 
			[\bx(\calS_{k}), \Tilde{\bA}]\hat{\bgamma}_k$
		\end{itemize}
		\item[4.] Sum of Root Mean Square Predictive Error (RMSPE) over $K$ folds
\begin{itemize}
\item Initialize $e = 0$\\
\text{ } for ($k$ in $1:K$) \\
\text{ } \quad \quad $ e =e + \sqrt{\sum_{j = 1}^q \sum_{\bs\ in \calS_k} (\by_j(\bs) - \hat{\by}_j(\bs))^2 / (|\calS_k|q)}$
\end{itemize}
\item[5.] Cross validation for choosing $\psi$ and $\alpha$
\begin{itemize}
\item Repeat steps (2) - (4) for all candidate values of $\psi$ and $\alpha$
\item Choose $\psi_0$ and $\alpha_0$ as the value that minimizes the sum of RMSPE
\end{itemize} 
	\end{enumerate}	
	\vspace*{-8pt}
	\rule{\textwidth}{1pt}
}}


\subsection{Comparison of Response and Latent Models}\label{subsec: response_vs_latent_NNGP}
Modeling the response as an NNGP produces a different model from modeling the latent process as an NNGP. In the former, Vecchia approximation to the joint density of the response yields a sparse precision matrix for the response. In the latter, it is the precision matrix of the realizations of the latent process that is sparse. This has been discussed in \cite{datta16} and also explored in greater generality by \cite{katzfuss2017general}. Comparisons based on the Kullback-Leibler divergence (KL-D) between the NNGP based models and their parent full GP models reveal that the latent NNGP model tends to be closer to the full GP than the response NNGP. A proof of such a result is provided by \cite{katzfuss2017general}, but this result holds only in the context of an augmented directed acyclical graphical model with nodes comprising the response and the latent variables. However, if we compute the KL-D between the NNGP models and their full GP counterparts in terms of the collapsed or marginal distribution for $\bY$, then it is theoretically possible for the response model to be closer to the full GP. 

Here we provide a simple example where a response NNGP model outperforms a latent NNGP model on a collapsed space. Consider $q=1$ variable. Assume that the observed location set is $\calS = \{\bs_1, \bs_2, \bs_3\}$, the latent process realization $\bomega(\calS)$ has covariance matrix $\sigma^2 \bR$ and $\by(\calS)$ has covariance matrix $\sigma^2\bR + \tau^2\bI_3$, where
\begin{equation}
\bR = \begin{bmatrix}
1& \rho_{12}  & \rho_{13}\\ 
\rho_{12}& 1 & \rho_{23} \\ 
\rho_{13}& \rho_{23} & 1
\end{bmatrix}\; .
\end{equation}
Let us construct the response NNGP and latent NNGP models using Vecchia's approximation in (\ref{eq: vecchia_approx}) with neighbor sets $N_m(\bs_2)=\{\bs_1\}$ and $N_m(\bs_3) = \{\bs_2\}$. 
Then the covariance matrix of $\by(\calS)$ from the response NNGP model and of $\by(\calS)$ from the latent NNGP model on the collapsed space, i.e., after $\bomega(\calS)$ is integrated out, are
\begin{equation}
\bSigma_R = \sigma^2 \begin{bmatrix}
1 + \delta^2 & \rho_{12}  & \frac{\rho_{12} \rho_{23}}{ 1 + \delta^2}\\ 
\rho_{12}& 1 + \delta^2 & \rho_{23} \\ 
\frac{\rho_{12} \rho_{23}}{ 1 + \delta^2}& \rho_{23} & 1 + \delta^2
\end{bmatrix}\; \mbox{ and } \;
\bSigma_l = \sigma^2 \begin{bmatrix}
1 + \delta^2 & \rho_{12}  & \rho_{12} \rho_{23}\\ 
\rho_{12}& 1 + \delta^2 & \rho_{23} \\ 
\rho_{12} \rho_{23}& \rho_{23} & 1 + \delta^2 
\end{bmatrix}\;,
\end{equation}
respectively, where $\delta^2 = \frac{\tau^2}{\sigma^2}$ is the noise-to-signal ratio with $\tau^2$ as the variance of the noise process $\epsilon(s)$. Since $\bR$ is positive-definite, we must have
\begin{equation}
1- (\rho_{12}^2 + \rho_{13}^2 + \rho_{23}^2) + 2\rho_{12}\rho_{13}\rho_{23} > 0 \; ,\; 1 - \rho_{12}^2 > 0\; .
\end{equation}
It is easy to show that $\bSigma_R$ and $\bSigma_l$ are also positive-definite. If $\rho_{13} = \frac{\rho_{12} \rho_{23}}{ 1 + \delta^2}$, then the KL-D from the response NNGP model to the true model always equals zero, which is no more than the KL-D from the latent NNGP model to the true model. If $\rho_{13} = \rho_{12} \rho_{23}$, then the KL-D of the latent NNGP model to the true model always equals zero, which reverses the relationship. Numerical examples can be found in \url{https://luzhangstat.github.io/notes/KL-D_com.html}

While theoretically one model does not always excel over the other, our simulations indicate that the latent NNGP model tends to outperform the response NNGP model in approximating their parent GP based models. This is consistent with the theoretical result of \citep{katzfuss2017general} and also with our intuition: the presence of the latent process should certainly improve the goodness of fit of the model. Without loss of generality, our discussion here considers the univariate case, but the argument applies to the multivariate setting as well. For the remainder of this section of the manuscript, let $\{y(\bs): \bs \in \mathcal{D}\}$ be the process of interest over $\mathcal{D} \subset \mathbb{R}^d, d \in N^+$, and let $y(\bs)  = \omega(\bs) + \epsilon(\bs)$ for some latent spatial GP $\omega(\bs)$ and white noise process $\epsilon(\bs)$. A response NNGP model specifies the NNGP on $y(\bs)$, while a latent NNGP model assumes that $\omega(\bs)$ follows the NNGP. The latter induces a spatial process on $y(\bs)$ too, but it is not an NNGP. 

Let the covariance matrix of $\by = y(\calS)$ of the parent GP based models be $\bC + \tau^2\bI$, where $\bC$ is the covariance matrix of the latent process $\omega(\calS)$. 
{Consider ${\tilde{\bC}}^{-1}$, the Vecchia approximation of the precision matrices $\bC^{-1}$, and $\tilde{\bK}^{-1}$, the Vecchia approximation of $\bK^{-1} = (\bC + \tau^2 \bI)^{-1}$.} 
The covariance matrix of $y(\calS)$ from the latent NNGP model is $\tilde{\bC} + \tau^2 \bI$, while the precision matrix of $y(\calS)$ from the response NNGP model is $\tilde{\bK}^{-1}$. We denote the error matrix of the Vecchia approximation of $\bC^{-1}$ by $\bE$. We assume that $\bE$ is small so that $\tilde{\bC}^{-1}$ approximates $\bC^{-1}$ well. With the same observed location $\calS$ and the fixed number of nearest neighbors, the error matrix of the Vecchia approximation of $\bK^{-1}$ is believed to be close to $\bE$, i.e.,
\begin{equation}
\bC^{-1} = \tilde{\bC}^{-1} + \bE\; ; \; \bK^{-1} = \tilde{\bK}^{-1} + \mathcal{O}(\bE).
\end{equation}
Representing the precision matrices of $y(\calS)$ of the parent GP based model and the latent NNGP model by
\begin{equation}
\begin{aligned}
(\bC+\tau^2\bI)^{-1} &= \bC^{-1} - \bC^{-1}\bM^{-1}\bC^{-1}\; , \bM = \bC^{-1} + \tau^{-2} \bI\; ,\\
(\tilde{\bC} + \tau^2\bI)^{-1} &= \tilde{\bC}^{-1} - \tilde{\bC}^{-1} \bM^{\ast-1}\tilde{\bC}^{-1}\; ,
\bM^\ast = \tilde{\bC}^{-1} + \tau^{-2}\bI\; ,
\end{aligned}
\end{equation}
we find that the difference between the precision metrics over the collapsed space for the parent NNGP and for the latent NNGP model is
\begin{align*}
&(\bC+\tau^2\bI)^{-1} - (\tilde{\bC}+ \tau^2\bI)^{-1} 
= \bC^{-1} - \bC^{-1}\bM^{-1}\bC^{-1} - \tilde{\bC}^{-1} + \tilde{\bC}^{-1} \bM^{\ast-1}\tilde{\bC}^{-1}\\
&\quad= \underbrace{\bE - \bE\bM^{-1}\tilde{\bC}^{-1} - \tilde{\bC}^{-1}\bM^{-1}\bE - \tilde{\bC}^{-1}(\bM^{-1} - \bM^{\ast-1})\tilde{\bC}^{-1}}_{\bB} - \underbrace{\bE\bM^{-1}\bE}_{\mathcal{O}(\bE^2)}
\end{align*}
Representing $\bB$ in terms of $\tilde{\bC}^{-1}$, $\bM^\ast$ and $\bE$, where $\bE$ is assumed to be nonsingular, we find
\begin{equation}
\label{eq: B_1}
\begin{aligned}
&\bB = \;\bE - \bE\bM^{\ast-1}\tilde{\bC}^{-1} + \bE\bM^{\ast-1}(\bE^{-1} + \bM^{\ast-1})^{-1}\bM^{\ast-1}\tilde{\bC}^{-1} - \tilde{\bC}^{-1}\bM^{\ast-1}\bE \\
&\quad+ \tilde{\bC}^{-1}\bM^{\ast-1}(\bE^{-1} + \bM^{\ast-1})^{-1}\bM^{\ast-1}\bE
+\tilde{\bC}^{-1}\bM^{\ast-1}(\bE^{-1} + \bM^{\ast-1})^{-1}\bM^{\ast-1}\tilde{\bC}^{-1}\; .
\end{aligned}
\end{equation}
Using the familiar Woodbury matrix identity and the expansion $(\bI + \bX)^{-1} = \sum_{n = 0}^{\infty} \{-\bX\}^{n}$, we find
\begin{align*}
(\bE^{-1} + \bM^{\ast-1})^{-1}\bM^{\ast-1} &= \{\bM^{\ast}(\bE^{-1} + \bM^{\ast-1})\}^{-1} = \{\bM^\ast \bE^{-1} + \bI\}^{-1} \\
& = \bI - \{\bI + \bE \bM^{\ast-1}\}^{-1} = \bI - \{\bI - \bE\bM^{\ast-1} + \mathcal{O}(\bE^2)\} \\
&= \bE\bM^{\ast-1} + \mathcal{O}(\bE^2)\;.
\end{align*}
Using the above equations and excluding the terms of order $\mathcal{O}(\bE^2)$ in the expression of $\bB$, the leading term in the difference is
\begin{equation}
\bB = (\bI - \tilde{\bC}^{-1}\bM^{\ast-1})\bE (I - \bM^{\ast-1}\tilde{\bC}^{-1})  =
(\bI + \tau^2 \tilde{\bC}^{-1})^{-1}\bE(\bI + \tau^2 \tilde{\bC}^{-1})^{-1} \; .
\end{equation}
Using the spectral decomposition $(\bI + \tau^2 \tilde{\bC}^{-1}) = \bP^\top (\bI + \tau^2 \bD)\bP$, where $\bP$ is orthogonal and
$\bD$ is diagonal with positive elements on the diagonal, we obtain
\begin{equation}
\begin{aligned}
\|\bB\|_{F} &= \|\bP^\top (\bI + \tau^2 \bD)^{-1}\bP\bE\bP^\top(\bI + \tau^2 \bD)^{-1}\bP\|_F = \| (\bI + \tau^2 \bD)^{-1}\bP\bE\bP^\top(\bI + \tau^2 \bD)^{-1}\|_F\\
 &\leq \| \bP\bE\bP^\top\|_F = \|\bE\|_F\; ,
\end{aligned}
\end{equation}
where $\|\cdot\|_F$ denotes the Frobenius matrix norm. The inequality also holds for the absolute value of the determinant and $p$ norms. And the equality holds if and only if $\tau^2 = 0$ when the difference is the same as the error matrix for response NNGP model. Thus, the latent model tends to shrink the error from the Vecchia approximation, which explains the expected superior performance of the latent NNGP model over the response NNGP model based on KL-Ds.

\section{Simulation}\label{sec: simulation}

We implemented our models in the \texttt{Julia}~1.2.0 numerical computing environment \citep{bezanson2017julia}. All computations were conducted on an Intel Core i7-7700K CPU @ 4.20GHz processor with 4 cores each and 2 threads per core---totaling 8 possible threads for use in parallel---and running a Linux Operating System (Ubuntu 18.04.2 LTS) with 32 Gbytes of random-access memory. Model diagnostics and other posterior summaries were implemented within the \texttt{Julia} and the \texttt{R}~3.6.1 statistical computing environment. 

We simulated $\by(\bs_i)$s using {\eqref{eq: conj_latent_model}} with $q = 2, p = 2$ over $n=1200$ randomly generated locations inside a unit square. The following describes this process. First, a design matrix $\bX$ is fixed with its first column of $1$'s and a single predictor generated from a standard normal distribution. We then generated $\bomega(\bs_i)$'s over these locations and fixed their values. Finally, we generated $\by(\bs_i)$s using {\eqref{eq: conj_latent_model}}. An exponential covariance function with decay $\phi$ was used to model $\rho_\psi(\cdot, \cdot)$ in {\eqref{eq: conj_latent_model}}, i.e., 
$
\rho_\psi(\bs', \bs'') = \exp{(-\phi\|\bs' - \bs''\|)}, \text{ for } \bs', \bs'' \in {\cal D}\;,
$
where $\|\bs' - \bs''\|$ is the Euclidean distance between $\bs'$ and $\bs''$, and $\psi = \phi$. The parameter values fixed to generate the data are listed in Table~\ref{table:sim1}. We withheld 200 locations to evaluate predictive performance for conjugate models and benchmark models. 

For our analysis, we assigned a flat prior for $\bbeta$, a prior of $\bSigma\sim \mbox{IW}(\bPsi, \nu)$ with $\bPsi = \bI_2$ (the $2\times 2$ identity matrix) and $\nu = 3$. 
{The candidate values for $\{\phi, \alpha\}$ in our cross-validation algorithms were chosen over a 25 by 25 grid defined by the range $[2.12, 26.52] \times [0.8, 0.99]$, where the support of $\phi$ corresponds to an effective spatial range (i.e., the distance where the spatial correlation drops to below 0.05) between $\sqrt{2}/12.5$ to $\sqrt{2}$, and the support of $\alpha$ is an interval centered at the actual value with width $0.2$. We used $K=5$ fold cross-validation for choosing the hyperparameters. This same setup was used for all the models to achieve a fairer comparison.} 

We used $500$ posterior samples for both the conjugate response and conjugate latent NNGP models. Note that these are draws from the exact posterior distribution, hence there are no issues of iterative convergence. The run times for the conjugate models included the time for choosing hyper-parameters through the cross-validation algorithm and the time for obtaining the posterior samples. Table~\ref{table:sim1} presents the posterior estimates of the regression coefficients $\bbeta = \{\bbeta_{ij}\}_{i = 1, j = 1}^{p, q}$, the covariance of the measurement error (labeled as $\mbox{cov}(\beps)$), covariance among the different latent processes (labeled as $\mbox{cov}(\bomega)$; this applies only to the latent NNGP model) and hyperparameters $\{\phi, \alpha\}$ in Table~\ref{table:sim1}.

\begin{table}[t]
\caption{Simulation study summary table: posterior mean (2.5\%, 97.5\%) percentiles}
\begin{minipage}[t]{\textwidth}
	\centering
	\begin{tabular}{c|c|cc}
	\hline\hline
			& True & Conj resp & Conj latent  \\
			\hline
			$\bbeta_{11}$ &  1.0 & 1.391 (0.814,1.902) & 1.459 (0.865,2.057) \\
			$\bbeta_{12}$ & 1.0 & 0.813 (0.344,1.286) &  0.734 (0.201,1.276) \\
			$\bbeta_{21}$ & -2.0 & -1.978 (-2.114,-1.841) &  -1.979 (-2.121,-1.842) \\
			$\bbeta_{22}$ &  2.0 &  2.076 (1.952,2.21) & 2.082 (1.961,2.208) \\
			$\mbox{cov}(\beps)_{11}$ &   0.222  &  0.226 (0.205,0.248) &  0.231 (0.212,0.252)\\
			$\mbox{cov}(\beps)_{12}$ & -0.111  & -0.113 (-0.129,-0.099) &  -0.115( -0.128,-0.103)\\
			$\mbox{cov}(\beps)_{22}$ &  0.167 & 0.172 (0.158,0.188) & 0.175 (0.16, 0.189) \\
			$\mbox{cov}(\bomega)_{11}$ & 1.234 & -- & 1.208 (1.148,1.268) \\
			$\mbox{cov}(\bomega)_{12}$ & -0.701 & -- & -0.705 ( -0.75,-0.658)\\
			$\mbox{cov}(\bomega)_{22}$ & 1.077 & -- &  1.077 (1.023,1.131)\\
			$\phi$& 6.0 & 8.220  & 7.204 \\
			$\alpha$& 0.9 &  0.863 & 0.871 \\
			\hline
			RMSPE &--& \footnotemark[1][0.727; 0.602; 0.668] &  \footnotemark[1][0.723; 0.6; 0.664]\\
			MSEL &--& -- & \footnotemark[1][0.112; 0.112; 0.103] \\
			CVG &--& \footnotemark[1][0.935; 0.955; 0.945] & \footnotemark[1][0.925; 0.95; 0.9375]\\
			CVGL &--& -- & \footnotemark[1][0.957; 0.945; 0.951]\\
			{MCRPS} &-- & \footnotemark[1][-0.408; -0.336; -0.372] & \footnotemark[1][-0.405; -0.334; -0.37]\\
			time(s) &--& \footnotemark[2][12; 1] & \footnotemark[2][17; 1] \\
			\hline\hline 
	\end{tabular}
\footnotetext[1]{[response 1; response 2; all responses]}
\footnotetext[2]{[time for cross-validation in seconds; time for sampling in seconds]}
\label{table:sim1}
\end{minipage}
\end{table}

Table~\ref{table:sim1} lists the parameter estimates and performance metrics of the candidate models. The NNGP models used in these experiments used $m = 10$ nearest neighbors. 
{The posterior inference of regression slopes $\{\bbeta_{21}, \bbeta_{22}\}$ are similar between the response and latent models.} The 95\% 
{credible} intervals of the intercepts $\{\bbeta_{11}, \bbeta_{12}\}$ include the value used to generate the data. The covariance matrix of the measurement errors $\mbox{cov}(\beps)$ is defined as $(\alpha^{-1}-1)\bSigma$ and computed using the posterior samples of $\bSigma$ and the value of $\alpha$ obtained from cross-validation. The posterior samples of $\mbox{cov}(\bomega)$ are computed directly from the posterior samples of $\bomega$. 
The conjugate NNGP models all yielded very similar RMSPEs {and MCRPSs}. The CVG and CVGL 
are close to 0.95, supporting reliable inference from conjugate NNGP models. 
{The run time required by both conjugate models are less than 20 seconds.}
The simulation example shows that fitting a conjugate model is a pragmatic method for quick inference in multivariate spatial data analysis. 

\begin{figure}[!ht]
     \subfloat[$\bomega_1 + \bbeta_{11}$ true\label{subfig:sim1a}]{%
       \includegraphics[width=0.45\textwidth]{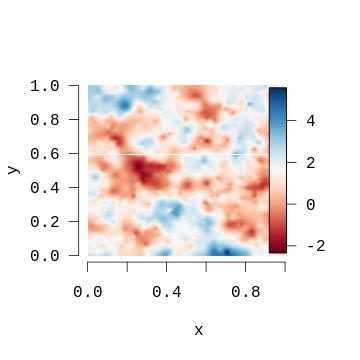}
     }
     \hfill
     \subfloat[$\bomega_1 + \bbeta_{11}$ latent NNGP\label{subfig:sim1b}]{%
       \includegraphics[width=0.45\textwidth]{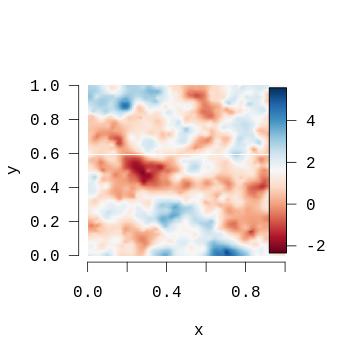}
     }\\
     
     \subfloat[$\bomega_2 + \bbeta_{12}$ true\label{subfig:sim1e}]{%
       \includegraphics[width=0.45\textwidth]{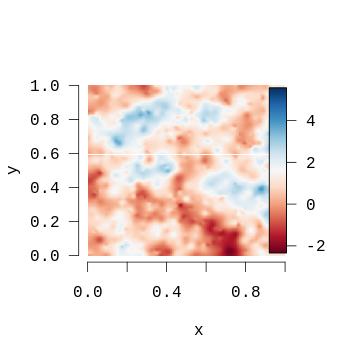}
     }
     \hfill
     \subfloat[$\bomega_2 + \bbeta_{12}$ latent NNGP\label{subfig:sim1f}]{%
       \includegraphics[width=0.45\textwidth]{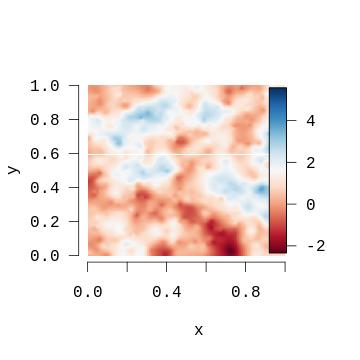}
     }
     \caption{Panels~(a)~and~(c) present interpolated maps of the true generated latent processes. Panels~(b)~and~(d) present the posterior means of the spatial latent processes $\bomega$ estimated using the conjugate latent NNGP model for the true processes (a)~and~(c), respectively. The NNGP based models were all fit using $m = 10$ nearest neighbors.\label{fig:sim1}}
   \end{figure}

Figure~\ref{fig:sim1} presents interpolated maps of the posterior means of the latent processes. Panel~(a) presents an interpolated map of the values of the first spatial process $\omega_1(\bs)$ added to the corresponding intercept $\beta_{11}$ over the unit square. Panel~(b) is the corresponding posterior estimate from the latent process model. Panels~(c)~and~(d) are the corresponding interpolated maps for the second process $\omega_2(\bs)$ and the corresponding estimate $\beta_{12}$. The similarity in spatial patterns between the two data generating processes and their corresponding estimates reveal that these models and their fitting algorithms are able to effectively capture the features of the underlying processes and the differences between them.  

\section{Normalized Vegetation Index Data Analysis}\label{sec: real_data_analy}
We implemented all our proposed models on a vegetation index and land cover data \citep[see][for further details]{ramon2010modis, sulla2018user}.
{We deal with two outcomes: (i) standard Normalized Difference Vegetation Index (NDVI); and (ii) red reflectance. NDVI is a robust and empirical measure of vegetation activity on the land surface that is important for understanding the global distribution of vegetation types, their biophysical and structural properties, and spatial-temporal variations \citep{ramon2010modis}. Red reflectance measures the spectral response in the red (0.6-0.7 $\mu m$) wavelengths region. Both outcomes are sensitive to the vegetation amount.
All data were 
mapped to Euclidean planar coordinates using the sinusoidal (SIN) grid projection following \cite{banerjee2005geodetic}. For the current analysis we restrict ouselves to zone \textit{h08v05}, which runs between 11,119,505 and 10,007,555 meters south of the prime meridian and between 3,335,852 to 4,447,802 meters north of the equator. This corresponds to the western United States. We included an intercept specific to each outcome and an indicator variable for no vegetation (or urban area) through the 2016 land cover data as our explanatory variables. All other data were measured through the MODIS satellite over a 16-day period from 2016.04.06 to 2016.04.21. Some variables were rescaled and transformed subsequent to some exploratory data analysis for numerical robustness. The data sets were downloaded using the \texttt{R} package \textit{MODIS}, and the code for exploratory data analysis is also available at \url{https://github.com/LuZhangstat/Conj_Multi_NNGP}.}  

\begin{figure}[!ht]
     \subfloat[ \label{subfig:real_conj_latent_mapsa}]{%
       \includegraphics[width=0.2\textwidth]{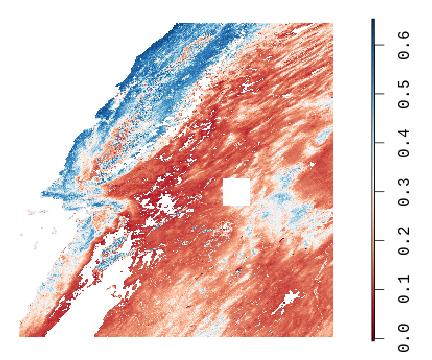}
     }
     \hfill
     \subfloat[ \label{subfig:real_conj_latent_mapsb}]{%
       \includegraphics[width=0.2\textwidth]{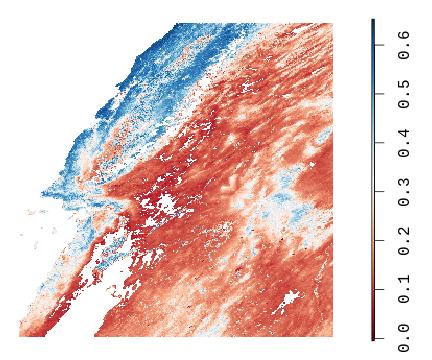}
     }
     \hfill
     \subfloat[ \label{subfig:real_conj_latent_mapsc}]{%
       \includegraphics[width=0.2\textwidth]{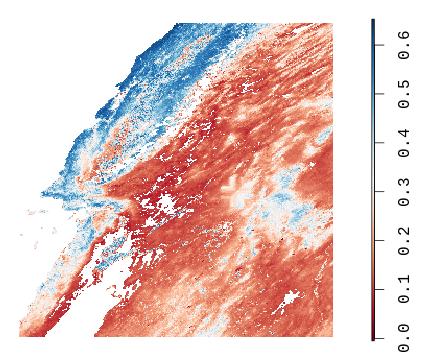}
     }
     \hfill
     \subfloat[ \label{subfig:real_conj_latent_mapsd}]{%
       \includegraphics[width=0.2\textwidth]{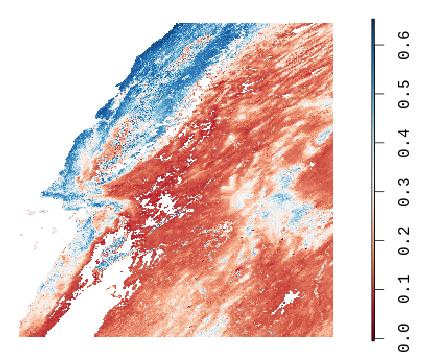}
     }\\
     
     \subfloat[ \label{subfig:real_conj_latent_mapse}]{%
       \includegraphics[width=0.2\textwidth]{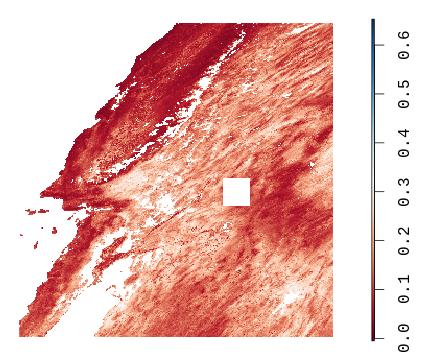}
     }
     \hfill
     \subfloat[\label{subfig:real_conj_latent_mapsf}]{%
       \includegraphics[width=0.2\textwidth]{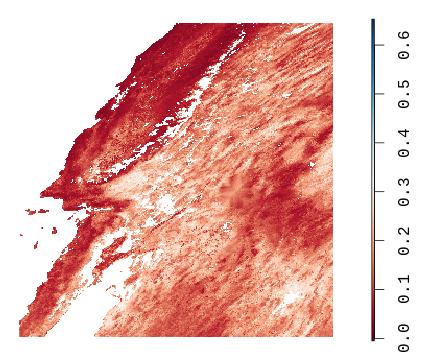}
     }
     \hfill
     \subfloat[\label{subfig:real_conj_latent_mapsg}]{%
       \includegraphics[width=0.2\textwidth]{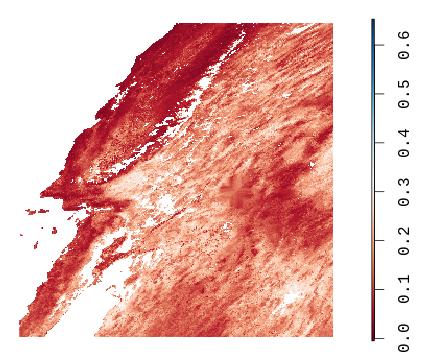}
     }
     \hfill
     \subfloat[\label{subfig:real_conj_latent_mapsh}]{%
       \includegraphics[width=0.2\textwidth]{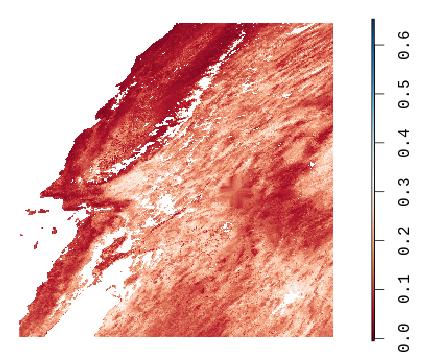}
     }
     \caption{Colored NDVI and red reflectance images (first and second rows, respectively) of the western United States (zone h08v05). Panels (a)~and~(e) presented interpolated maps of the raw data, (b)~and~(f) present interpolated posterior predictive means of NDVI and red reflectance from the conjugate NNGP response models, (c)~and~(g) present the posterior predictive maps of NDVI and red reflectance from the conjugate NNGP latent models, and (d)~and~(h) present the posterior predictive means of the intercept-centered latent processes corresponding to NDVI and red reflectance recovered from conjugate NNGP latent model.
\label{fig:real_conj_latent_maps}}
\end{figure}

There are 3,115,934 observed locations. We used a transformed NDVI ($\log(\mbox{NDVI} + 1)$ labeled as NDVI) and red reflectance (red reflectance) as responses. The NNGP based models were constructed with $m = 10$ nearest neighbors.
{We held out NDVI and red reflectance on 67,132 locations 
with about half of them in the region between 10,400,000 and 10,300,000 meters south of the prime meridian and between 3,800,000 and 3,900,000 meters north of the equator. We evaluate the predictive performance of our models using these held out locations and use the remaining for training the models. 
Figure~\ref{subfig:real_conj_latent_mapsa} illustrates the map of the transformed NDVI data. The white square is the region held out for prediction.} 

Posterior inference from our conjugate models were based on 500 independent samples drawn directly from the exact posterior distribution. Since these samples are directly drawn from the conjugate posterior distribution, there is no need to monitor convergence of these samples. 
{We assigned a flat prior for all regression coefficients, while we assumed $\bSigma\sim \mbox{IW}(\bPsi, \nu)$ with $\bPsi = \mbox{diag}([1.0, 1.0])$ and $\nu = 3$.} We recursively shrink the domain and grid of candidate values $\{\phi, \alpha\}$ through repeatedly using the cross-validation algorithm in Section~\ref{subsec: CV_conj_NNGP} for selecting and fixing these parameters. The recorded run time for running the cross-validation algorithms, therefore, varied substantially across different models. 
{We took $K=5$ folds in our cross-validation algorithm and ran the prediction on each fold in parallel}. The subsequent computing of all the code were run with a single thread.

The results for the conjugate models are listed in Table~\ref{table:real_conj}. 
{Note that the cross-validation yielded optimal values of $\alpha \approx 1$, which means a negligible nugget. This causes the predictions on the hold-out sets to be smooth.} Consistent with the related background, the regression coefficients of the index of ``no vegetation'' (urban area) are significantly negative for NDVI, which is to be expected as NDVI is a measure of greenness and low vegetation represents lack of greenness. On the other hand, ``no vegetation'' (urban area) is significantly positive for red reflectance, which is also consistent with the fact that NDVI and red reflectance tend to be negatively associated. In fact, this negative association is seen to persist even after we account for the ``no vegetation'' index as seen from the estimated covariance matrices for the residual noise in both models and the latent spatial process for the latent NNGP model.

Model performances were compared in terms of RMSPE, CVG, {M}CRPS and run time. The spatial models, unsurprisingly, greatly improved predictive accuracy. In fact, conjugate Bayesian spatial models effected a 35\% shrinkage in the magnitude of RMSPE over a non-spatial Bayesian linear model, i.e., with $\bomega(\bs)=\mathbf{0}$ in (\ref{eq: conj_latent_model}). Therefore, we do not show the estimates from the non-spatial model. 
Table~\ref{table:real_conj} shows that the latent model seems to slightly outperform the response model in terms of RMSPE and MCRPS, while the CVG for both these models are very comparable. Posterior sampling for the conjugate response and latent models cost between 1.8 and 18.88 minutes, respectively, which is impressive given our sample sizes of around $3\times 10^6$ locations. The run time for the cross-validation algorithm and the posterior sampling from the conjugate models is appealing for such massive data sets. 

\begin{table}[!t]
\caption{Vegetation Index data analysis: Posterior means (2.5\%,97.5\%) percentiles, model comparison metrics and run times (in minutes).}
\centering
\begin{minipage}[t]{\textwidth} 
	\begin{tabular}{ccccc}
	\hline\hline
			& conj response & conj latent \\
			\hline
			$\mbox{intercept}_1$ & 0.1023 (0.0822,0.1223)& 0.240729 (0.240723,0.240736) \\
			$\mbox{intercept}_2$ & 0.2218 (0.2094,0.2338)& 0.144277 (0.144273,0.144281)  \\
			$\mbox{no vege or urban area}_1$ & -8.010e-3 (-8.233e-3,-7.796e-3) & -8.025e-3 (-8.050e-3,-8.001e-3) \\
			$\mbox{no vege or urban area}_2$ & 4.381e-3 (4.261e-3,4.514e-3) & 4.390e-3 (4.376e-3,4.402e-3)\\
			$\mbox{cov}(\beps)_{11}$ & 3.493e-5 (3.487e-5,3.499e-5) & 3.125e-5 (3.120e-5,3.130e-5)\\
			$\mbox{cov}(\beps)_{12}$ & -1.214e-5 (-1.217e-5,-1.212e-5) & -1.086e-5 (-1.089e-5,-1.085e-5)\\
			$\mbox{cov}(\beps)_{22}$ &1.090e-5 (1.089e-5,1.092e-5) & 9.760e-6 (9.745e-6,9.776e-6)\\
			$\mbox{cov}(\bomega)_{11}$ & -- & 1.7192e-2 ( 1.7190e-2,1.7193e-2) \\
			$\mbox{cov}(\bomega)_{12}$ & -- & -7.0307e-3 (-7.0314e-3,-7.03e-3) \\
			$\mbox{cov}(\bomega)_{22}$ & -- & 3.8897e-3 (3.8893e-3,3.8901e-3)\\
			$(\phi,\alpha)$ & (17.919,0.999551) &  (20.1755,0.999551)\\
			\hline
			RMSPE & \footnotemark[1][0.05707;  0.03187;  0.04622] &  \footnotemark[1][0.0503;  0.02572;  0.03995]\\
			{MCRPS} & \footnotemark[1][-0.03301;  -0.0188;  -0.02591]& \footnotemark[1][-0.0314;  -0.01748;  -0.02444]\\
			CVG & \footnotemark[1][0.9756; 0.9707; 0.9732] & \footnotemark[1][0.9764; 0.9715; 0.974]\\
			time(mins) & \footnotemark[2][1012.18; 1.8] & \footnotemark[2][270.28; 18.88] \\
			\hline\hline
	\end{tabular}
\footnotetext[1]{[response 1; response 2; all responses]}
\footnotetext[2]{[time for cross-validation in minutes; time for generating 500 samples in minutes]}
\end{minipage}
\label{table:real_conj}
\end{table}

Visual inspections of the predictive surfaces based on the conjugate response NNGP model are depicted in Figure~\ref{fig:real_conj_latent_maps}. The maps of the latent processes recovered by the conjugate latent NNGP shown in Figure~\ref{fig:real_conj_latent_maps} further corroborate the findings in Table~\ref{table:real_conj} regarding the negative association between the two latent processes for transformed NDVI and red reflectance. We see from the above figure that the blue and red regions for NDVI seem to have been swapped in the map for red reflectance. Notably, the proposed methods smooth out the predictions in the held-out region which is also a consequence of the cross-validation estimate of $\alpha\approx 1$.  

\section{Summary and Discussion}\label{sec: Conclusions}
We have presented a conjugate Bayesian multivariate spatial regression model using Matrix-Normal and Inverse-Wishart distributions. A specific contribution is to embed the latent spatial process within an augmented Bayesian multivariate regression to obtain posterior inference for the high-dimensional latent process with stochastic uncertainty quantification. For scalability to massive spatial datasets---our examples here comprise locations in the millions---we adopt the increasingly popular Vecchia approximation and, more specifically, the NNGP models that render savings in terms of storage and floating point operations. We present elaborate simulation experiments to test the performance of different models using datasets exhibiting different behaviors. Our conjugate modeling framework fixes hyperparameters using a $K$-fold cross-validation approach. While our analysis is based upon fixing these hyperparameters, the subsequent inference obtained is seen to be effective in capturing the features of the generating latent process (in our simulation experiments) and is orders of magnitude faster than iterative alternatives at such massive scales as ours. We also applied our models, and compared them, in our analysis of an NDVI dataset. The scalability of our approach is guaranteed when univariate scalable model can exploit a tractable precision or covariance matrix. Our approach can, therefore, incorporate other methods such as multiresolution approximation (MRA) and more general Vecchia-type of approximations \cite[see, e.g.][]{katzfuss2017general,pbf2020}. 
 
Future work can extend and adapt this framework to univariate and multivariate spatiotemporal modeling. A modification is to use a dynamic nearest-neighbor Gaussian process (DNNGP) \citep{datta16b} instead of the NNGP in our models, which dynamically learns about space-time neighbors rather than fixing them. We can also develop conjugate Bayesian modeling frameworks for spatially-varying coefficient models, where the regression coefficients $\bbeta$ are themselves random fields capturing the spatially-varying impact of predictors on the vector of outcomes. While conceptually straighforward, their actual implementation at massive scales will require substantial development. 

Developments in scalable statistical models must be accompanied by explorations in high performance computing. While the algorithms presented here are efficient in terms of storage and flops, they have been implemented on modest hardware. Implementations exploiting Graphical Processing Units (GPUs) and parallel CPUs can be further explored. For the latent NNGP models, the algorithms relied upon sparse solvers such as conjugate gradients and LSMR matrix algorithms. Adapting such libraries to GPUs and other high performance computing hardware will need to be explored and tested further in the context of our spatial Gaussian process models.

\section*{Supporting Information}
The work of the first and second authors was supported, in part, by
federal grants NSF/DMS 1513654, NSF/IIS 1562303, and NIH/NIEHS
1R01ES027027. The third author was supported by NSF/EF 1253225 and
NSF/DMS 1916395, and National Aeronautics and Space Administration's
Carbon Monitoring System project.

\appendix

\bibliographystyle{ba}  
\bibliography{lubib} 

\label{lastpage}

\end{document}